\documentclass[10pt,aps,showpacs,nofootinbib]{revtex4}

\usepackage{graphicx}
\usepackage{graphics}
\usepackage{xspace}
\usepackage{amssymb}
\usepackage{amsmath}
\usepackage{latexsym}
\usepackage{mathrsfs}

\newcommand{\half}{\textstyle{\frac{1}{2}}}

\newcommand{\g}{\gamma}

\newcommand{\be}{\begin{equation}}
\newcommand{\ee}{\end{equation}}
\newcommand{\bea}{\begin{eqnarray}}
\newcommand{\eea}{\end{eqnarray}}
\newcommand{\nn}{\nonumber}
\newcommand{\nslash}{\kern 0.2 em n\kern -0.50em /}
\newcommand{\kslash}{\kern 0.2 em k\kern -0.45em /}
\newcommand{\pslash}{\kern 0.2 em p\kern -0.50em /}
\newcommand{\Sslash}{\kern 0.2 em S\kern -0.50em /}
\newcommand{\Pslash}{\kern 0.2 em P\kern -0.50em /}
\newcommand{\Rslash}{\kern 0.2 em R\kern -0.50em /}
\newcommand{\Dslash}{\kern 0.2 em D\kern -0.65em /}
\newcommand{\open}{{<\kern -0.3 em{\scriptscriptstyle )}}}
\newcommand{\rr}{\vec{R}_T}

\newcommand{\eps}{\epsilon}
\newcommand{\ii}{{\rm i}}
\newcommand{\xbj}{x}
\newcommand{\de}{d}
\newcommand{\st}{T}
\newcommand{\tperp}{{T \perp}}
\newcommand{\tperpb}{{T}}
\newcommand{\sumint}{\kern 0.2 em {\textstyle\sum} \kern -1.1 em \int_X}

\allowdisplaybreaks[2]

\begin{document}
\title{
Two-hadron semi-inclusive production including subleading twist
}

\author{Alessandro Bacchetta}
\email{alessandro.bacchetta@physik.uni-regensburg.de}
\affiliation{Institut f\"ur Theoretische Physik, Universit\"at Regensburg,
D-93040 Regensburg, Germany}

\author{Marco Radici}
\email{radici@pv.infn.it}
\affiliation{Dipartimento di Fisica Nucleare e Teorica, 
Universit\`{a} di Pavia, 
and\\
Istituto Nazionale di Fisica Nucleare, Sezione di Pavia, I-27100 Pavia, Italy}

%%%%%%%%%%%%%%%%%%%%%%%%%%%%%%%%%%%%%%%%%%%%%%%%%%%%%%%%%%%%
\begin{abstract}
We extend the analysis of two-hadron fragmentation functions to the subleading
twist, discussing also the issue of color gauge invariance.
Our results can be used anywhere two unpolarized hadrons are semi-inclusively produced in 
the same fragmentation region, also at moderate values of the hard scale 
$Q$. Here, we consider the example of polarized 
deep-inelastic production of two hadrons and we give a complete list of cross sections and 
spin asymmetries up to subleading twist. Among the results, we highlight the 
possibility of extracting the transversity distribution with longitudinally 
polarized targets and also the twist-3 distribution $e(x)$, which is related to the 
pion-nucleon $\sigma$ term and to the strangeness content of the nucleon.
\end{abstract}


\maketitle

%%%%%%%%%%%%%%%%%%%%%%%%%%%%%%%%%%%%%%%%%%%%%%%%%%%%%%%%%%%%
\section{Introduction}
\label{sec:intro}

The study of the distribution of hadrons produced in the fragmentation of a quark
offers the opportunity to understand the mechanism of hadronization as well as to extract
information about the partonic structure of hadrons; both issues are a manifestation of 
confinement in QCD, a yet unexplained phenomenon. So far, parametrizations are available only for
the distribution of the longitudinal momentum of only one of the final-state hadrons, the 
familiar unpolarized fragmentation function $D_1(z)$~\cite{Kretzer:2001pz,Kniehl:2000fe}. 
Clearly, most of the complexity of the fragmentation process lies unexplored.

When the transverse momentum of one of the outgoing hadrons is measured, a new
fragmentation function can be introduced relating the transverse polarization of 
the parent quark to the distribution of the produced hadron in the 
transverse direction~\cite{Collins:1993kk}. This so-called Collins function acts as an analyzing 
power and it is perhaps the simplest observable that reveals the role of the 
quark's spin in the hadronization process. It also acts as a filter to measure the still unknown
distribution of transverse spin of quarks (transversity, for a review see 
Ref.~\cite{Barone:2001sp}) and the tensor 
charge of the hadron thereof~\cite{Collins:1993kk}. However, the price to pay is the complete 
knowledge of the transverse dynamics of the detected 
leading hadron inside the jet. This creates problems both experimentally, as it is evident, and 
also theoretically, because the introduced dependence upon an intrinsic (nonperturbative) 
transverse momentum complicates the treatment of color gauge 
invariance~\cite{Boer:2003cm,Metz:2002iz} and evolution
equations~\cite{Henneman:2001ev,Kundu:2001pk,Boer:2001he,Goeke:2003az}. 

When two final-state hadrons are measured, in principle the number of variables doubles. 
For instance, it is possible to measure the relative transverse momentum of the pair, as 
well as its center-of-mass transverse momentum. Therefore, even after integrating upon the 
center-of-mass transverse momentum, a transverse vector is still available to 
establish a relation with the transverse polarization of the fragmenting
quark~\cite{Collins:1994ax,Jaffe:1998hf,Bianconi:1999cd}.

Already from this intuitive discussion it is evident that two-hadron
fragmentation functions can be important in studying spin
effects in hadronization. They are perhaps more challenging to measure,
inasmuch as they require the simultaneous detection of two hadrons inside the same 
jet. On the other side, the integration upon the center-of-mass transverse momentum
removes the above mentioned difficulty about the evolution equations and it avoids, 
at least at leading twist, the potential loss of universality implied by a correct 
treatment of color gauge invariance~\cite{Boer:2003cm,Metz:2002iz}, as it will also 
be discussed in Sec.~\ref{sec:correlator}. 

Another class of functions that deserves much attention is that of
polarized fragmentation functions. In this case, the spin of the final-state
hadron is measured and its relation with the hadronization dynamics can be
investigated. However, in general the spin of a final-state hadron can be
analyzed only through the decay into two or more hadronic byproducts. In this
sense, polarized fragmentation functions can be thought of as specific
examples of multi-hadron fragmentation functions. For instance, the
polarization of a vector meson  (e.g., $\rho^0$) is reflected in the angular
distribution of its decay products (e.g., $\pi^+ \pi^-$). As a consequence,
spin-1 polarized fragmentation functions~\cite{Ji:1994vw,Efremov:1982sh,Bacchetta:2000jk} 
correspond to the relative $p$-wave part of
two-hadron fragmentation functions~\cite{Bacchetta:2002ux}. 
At present, however, the formalism of two-hadron
fragmentation functions cannot comprise parity-violating decays, such as the
extremely important case of the $\Lambda$ 
baryon~\cite{Ji:1994vw,Artru:1990zv,Jaffe:1996wp,deFlorian:1998zj,Anselmino:2001ps,Anselmino:2001xc}.

Two-hadron fragmentation functions were first introduced in 
Ref.~\cite{Konishi:1978yx}, but with no quark polarization. Extension of the 
original functions to include polarization effects (usually known as interference fragmentation
functions) were studied in Refs.~\cite{Collins:1994kq,Collins:1994ax,Jaffe:1998hf,Artru:1996zu}.
The complete leading-twist analysis has been carried out in 
Ref.~\cite{Bianconi:1999cd} and employed in semi-inclusive 
DIS~\cite{Bianconi:1999uc,Radici:2001na} and electron-positron
annihilation~\cite{Boer:2003ya}. Positivity bounds and the expansion in the 
partial wave of the two hadrons were presented in Ref.~\cite{Bacchetta:2002ux}.
Very recently, a study of collinear fragmentation into two hadrons has been 
performed~\cite{deFlorian:2003cg}, demonstrating the factorization of two-hadron 
fragmentation functions at next-to-leading order in $\alpha_S$ and calculating their 
evolution, originally studied in Ref.~\cite{Vendramin:1981te}. 
In this article, we are going to extend the existing treatment to the
subleading-twist level, but integrating upon the transverse momentum. The way we
proceed is very similar to what was done in Ref.~\cite{Mulders:1996dh}, for
one-hadron production (see also Ref.~\cite{Jaffe:1993xb}), and in Ref.~\cite{Boer:2003cm}, 
for the issue of color gauge invariance of the quark-quark correlator, even though we will 
only present results integrated upon the transverse momentum. The extension to the subleading 
twist is an important step not only from a formal point of view, but also because the 
measurement of two-hadron leptoproduction can be attempted in experiments at
moderate $Q^2$, where subleading-twist contributions should not be neglected.

The paper is organized as follows. In Sec.~\ref{sec:kin}, we will briefly review the
kinematics for the semi-inclusive production of two unpolarized hadrons inside the
same current jet. In Sec.~\ref{sec:correlator}, we present the complete twist analysis up to
subleading order of the quark-quark and quark-gluon-quark fragmenting correlators,
discussing also the issue of color gauge invariance and of partial-wave expansion. In 
Sec.~\ref{sec:hadrontensor}, the
explicit expression of the hadronic tensor for the semi-inclusive production of two
unpolarized hadrons in Deep-Inelastic Scattering (DIS) is shown, including leading-
and subleading-twist contributions. In Sec.~\ref{sec:cross}, the corresponding cross
sections and spin asymmetries are discussed for different polarization states of the beam
and the target. Finally, in Sec.~\ref{sec:end} some conclusions are drawn.

%%%%%%%%%%%%%%% Fig. 1 %%%%%%%%%%%%%%%%%%%

\begin{figure}[h]
\begin{center}
\includegraphics[width=5 cm]{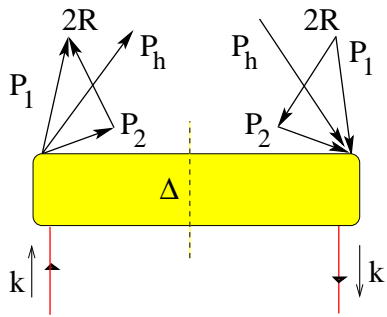}
\end{center}
\caption {Quark-quark correlation function $\Delta$ for the fragmentation of a
quark with momentum $k$ into a pair of  hadrons with total momentum $P_h=P_1+P_2$ 
and relative momentum $R=(P_1-P_2)/2$.}
\label{fig:fig1}
\end{figure}
%%%%%%%%%%%%%%%%%%%%%%%%%%%%%%%%%%%%%%%%%%%%%%%%%%%%%%%%%%%%%%%%%%%%%%%%%%%%%%

%%%%%%%%%%%%%%%%%%%%%%%%%%%%%%%%%%%%%%%%%%%%%%%%%%%%%%%%%%%%

\section{Kinematics}
\label{sec:kin}

The fragmentation process is schematically represented in Fig.~\ref{fig:fig1}, where a 
quark with mass $m$ and momentum $k$ fragments into two unpolarized hadrons with 
masses $M_1,\, M_2,$ and momenta $P_1,\, P_2$. We introduce the vectors $P_h=P_1+P_2$ and 
$R=(P_1-P_2)/2$. Using two dimensionless light-like vectors $n_+$ and $n_-$
(satisfying $n_+^2=n_-^2=0$ and $n_+\cdot n_-=1$), we describe a 4-vector $a$ as 
$[a^-,a^+,\vec a^{}_\st ]$ in terms of its light-cone components $a^\pm = a\cdot 
n_\mp =(a^0 \pm a^3)/\sqrt{2}$ and a bidimensional vector $\vec a^{}_\st$. 

For later use, transverse projection operators can be defined as 
\begin{equation} \begin{split} 
g_\st^{\mu \nu} &= g^{\mu \nu} - n_+^{\left\{ \mu \right.} n_-^{\left. \nu
\right\}} \\
\eps_\st^{\mu \nu} & \equiv \eps^{\rho \sigma \mu \nu} n_{ + \rho}\, n_{- \sigma}
\; ,
\end{split}
\label{eq:Tproj}
\end{equation} 
where the braces indicate symmetrization upon the included indices and
$\eps^{0123}=1$. In general, the fragmentation process is described in the frame 
where the hadronic final state has no transverse component, i.e., the frame where
$\vec P^{}_{h\,\st}=0$. We will see in the following that in actual measurements the
natural choice is different and the required boost introduces effects that have to
be consistently taken into account when extending the analysis to the subleading
twist. 

We also define the variables $z= P_h^-/k^-$, the light-cone fraction of 
fragmenting quark momentum carried by the hadron pair, and the variable $\zeta = 
2R^-/P_h^-$, which describes how the total momentum of the pair is split into the 
two single hadrons. Therefore, the relevant momenta can be parametrized as 
\begin{equation} \begin{split} 
   k^\mu &= \left[ \frac{P_h^-}{z}, \frac{z(k^2+\vec k_\st^{\, 2})}{2P_h^-},
   \vec k^{}_\st \right]  \\
   P_h^\mu &= \left[ P_h^-, \frac{M_h^2}{2 P_h^-}, \vec 0 \right]  \\
   R^\mu &= \left[ \frac{\zeta}{2} P_h^-,
    \frac{(M_1^2-M_2^2)-\textstyle{\frac{\zeta}{2}}M_h^2}{2P_h^-}, \vec R^{}_\st
\right] \; ,
\label{eq:kin}
\end{split} \end{equation}  
where $M_h$ is the pair invariant mass. Not all components of the 4-vectors are 
independent. In particular, we note that 
\begin{equation} \begin{split} 
   R^2 &= \frac{M_1^2+M_2^2}{2}\, - \, \frac{M_h^2}{4}  \\
   \rr^2 &= \frac{1}{2}\, \left[ \frac{(1-\zeta)(1+\zeta)}{2} M_h^2 - (1-\zeta)
     M_1^2 - (1+\zeta) M_2^2 \right]  \\
   P_h \cdot R &= \frac{M_1^2-M_2^2}{2}  \\
   P_h \cdot k &= \frac{M_h^2}{2z} + z\, \frac{k^2+\vec k_\st^{\, 2}}{2}  \\
   R \cdot k &= \frac{(M_1^2-M_2^2)-\textstyle{\frac{\zeta}{2}} M_h^2}{2z} + 
     z\zeta \, \frac{k^2+\vec k_\st^{\, 2}}{4} - \vec k^{}_\st \cdot \vec R^{}_\st \; .
\label{eq:kin2}
\end{split} \end{equation}  
The positivity requirement $\rr^2 \ge 0$ imposes the further constraint
\begin{equation} 
    M_h^2 \ge \frac{2}{1+\zeta} M_1^2 + \frac{2}{1-\zeta} M_2^2  \; .
\label{eq:mh}
\end{equation} 

Note that to avoid the introduction of a new hard scale in the process, all
invariants listed above have to be small compared to the hard scale $Q$ of the process 
(where $Q^2=-q^2$, with $q$ the momentum transfer).

%%%%%%%%%%%%%%%%%%%%%%%%%%%%%%%%%%%%%%%%%%%%%%%

\section{The fragmentation correlator up to subleading twist}
\label{sec:correlator}

The soft processes underlying the fragmentation are symbolically represented
by the  shaded blob in Fig.~\ref{fig:fig1} and are described in terms of hadronic 
matrix elements of nonlocal quark operators as 
\begin{equation}
  \Delta(k,P_h,R)= \sumint \; 
  \int \frac{\de^{4\!}\xi}{(2\pi)^4} \; e^{\ii k\cdot\xi}\;
  \langle 0|\, \psi(\xi) \,|P_h,R; X\rangle
  \langle X; P_h,R|\, \overline{\psi}(0)\,|0\rangle \; ,
  \label{eq:defDelta}
\end{equation}
where $\psi$ is the quark field operator. The above correlator is not color gauge invariant, 
as the quark fields are evaluated at two light-front separated space-time points, $0$ and 
$\xi$. To restore color gauge invariance, the so-called gauge link operator must be included, 
\begin{equation}
U_{[0,\xi]} = {\cal P} \exp \left( -\ii g \int_0^\xi dw \cdot A(w) \right) \; , 
\label{eq:deflink}
\end{equation}
where $A$ is the gluon field with coupling constant $g$, and ${\cal P}$ indicates a 
path-ordered exponential. It symbolically corresponds to attach all possible soft gluon 
lines to the soft blob of Fig.~\ref{fig:fig1} and resum their contribution. As such,
the corresponding diagrams will still be considered as tree-level contributions, since
the coupling $g$ can be reabsorbed in the definition of the correlator itself.  

The quark line in the fragmentation correlator has to come from a hard process that 
determines a dominant light-like direction. For the final state in a semi-inclusive 
DIS process, the hard scale $Q$ selects the $n_-$ direction as the dominant one with respect to 
the transverse and $n_+$ ones, which are suppressed as ${\cal O}(M_h)$ and ${\cal O}(1/Q)$, 
respectively. The integration upon the suppressed $n_+$ components of the momenta can be 
performed up to ${\cal O}(1/Q)$ leading to
\begin{equation}
  \Delta(z,\vec k_\st ,R)= \sumint \; 
  \int \frac{\de \xi^+ d\vec \xi_\st}{(2\pi)^3} \; e^{\ii k\cdot\xi}\;\left. 
  \langle 0|\, \psi(\xi) \,|P_h,R; X\rangle
  \langle X; P_h,R|\, \overline{\psi}(0)\,|0\rangle \right\vert_{\xi^- = 0}\; .
  \label{eq:defDeltaz}
\end{equation}
Similarly, in the gauge link it was usually assumed that the $A^+$ component of the gluon field 
is suppressed and, by neglecting the $\vec A_\st^{}$ component and by 
imposing the choice $A^-=0$ (the so-called light-cone gauge), the gauge link was reduced to unity. 
Recently, the problem of the evaluation of
such operator and of the gauge-invariant description of the quark-quark correlator  
has been studied in Refs.~\cite{Ji:2002aa,Belitsky:2002sm,Boer:2003cm}. We will 
address it following the analysis of Ref.~\cite{Boer:2003cm} for the case of
semi-inclusive DIS; the results can be easily generalized to the case of 
electron-positron annihilation. The proof in Ref.~\cite{Boer:2003cm} relies on 
counting the powers in $1/Q$ of the product of the fragmenting quark propagator and of 
the various components of the gluon fields attached to the soft blob, retaining only 
the leading and subleading contributions; these arguments are independent of the 
hadronic final state, and they are valid also for the two-hadron production, as long 
as the additional vector $R$ does not introduce any new hard scale.

%%%%%%%%%%%%%%% Fig. 2 %%%%%%%%%%%%%%%%%%%

\begin{figure}[h]
\begin{center}
\includegraphics[width=4.5cm]{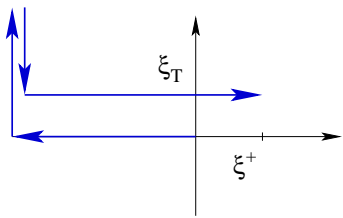}
\end{center}
\caption {Link structure for the leading-twist color gauge invariant quark-quark 
correlator for the fragmentation of a quark into a pair of hadrons.}
\label{fig:fig2}
\end{figure}

%%%%%%%%%%%%%%%%%%%%%%%%%%%%%%%%%%%%%%%%%%%%%%%%%%%%%%%%%%%%

It turns out that only a combined analysis of leading and subleading contributions
involving both $A^-$ and $\vec A_\st^{}$ components of the gluon field leads to a
color gauge invariant expression for the fragmentation correlator. Following 
Ref.~\cite{Boer:2003cm} (and generalizing its notation to the case of two-hadron 
production), a color gauge invariant object is obtained at leading twist by connecting 
the $0$ and $\xi$ points along the $+$ direction running through $-\infty$ and through 
the transverse directions at $\xi^+ = -\infty$ (see Fig.~\ref{fig:fig2} for a schematic 
picture of the link path), namely  
\begin{equation}
  \Delta^{[-]} (z,\vec k_\st ,R)= \sumint \; 
  \int \frac{\de \xi^+ \de \vec \xi_\st}{(2\pi)^3} \; e^{\ii k\cdot\xi}\; \left. 
  \langle 0| U^+_{[-\infty,\xi]} \, U^T_{[\infty,\xi]} \, \psi(\xi) \,|P_h,R; X\rangle
  \langle X; P_h,R|\, \overline{\psi}(0)\, U^T_{[0,\infty]} \, U^+_{[0,-\infty]} |0
  \rangle \right\vert_{\xi^- =0} \; ,
  \label{eq:linkDeltaz}
\end{equation}
where
\begin{equation} \begin{split} 
U^+_{[a,b]} &= {\cal P} \exp \left( \left. -\ii g \int_a^b dw^+ A^-(w) 
\right\vert_{
    \begin{array}{ccc} 
    \scriptstyle{w^- =}& \scriptstyle{b^- =}& \scriptstyle{a^-} \\ 
    \scriptstyle{\vec w^{}_\st =}& \scriptstyle{\vec b^{}_\st =}& 
    \scriptstyle{\vec a^{}_\st} \end{array}} \right)  \\
U^T_{[a,b]} &= {\cal P} \exp \left( \left. -\ii g \int_a^b d\vec w^{}_\st \cdot 
   \vec A^{}_\st(w) \right\vert_{
    \begin{array}{ccc} \scriptstyle{w^+ =}& \scriptstyle{b^+ =}& 
    \scriptstyle{a^+ } \\ 
    \scriptstyle{w^- =}& \scriptstyle{b^- =}& \scriptstyle{a^-} \end{array}} 
    \right) \; .
\label{eq:links}
\end{split} \end{equation}  
Note that by reabsorbing the product of gauge links, $U^T U^+$, into a redefinition
of the quark field $\psi$, the quark-quark correlator of Eq.~(\ref{eq:linkDeltaz}) 
falls back into the expression of Eq.~(\ref{eq:defDeltaz}), but for the $[-]$
superscript specifying the gauge link direction. Therefore, it still leads to a 
semipositive definite matrix in Dirac space~\cite{Bacchetta:2000kz} and 
the probabilistic interpretation of its leading-twist projections can be retained. The
dependence on the direction of the gauge link is due to the contribution of $U^T$, i.e.
of the transverse component of the gluon field at $\xi^+=-\infty$, which
plays a crucial role in $T$-odd effects since it introduces nontrivial phases in the
scattering amplitude. The direction of the gauge link depends on the considered
process, potentially posing a threat to the universality of the definition of the
soft correlator. For example, when considering the $e^+$-$e^-$ annihilation into one
pair of hadrons in the same jet the correlator of Eq.~(\ref{eq:linkDeltaz}) will depend 
on a gauge link running through $\xi^+ = +\infty$, therefore displaying the $[+]$ 
superscript. However, it has been explicitly shown that universality is preserved at 
the one-loop level~\cite{Metz:2002iz}.
%% Sulla frase seguente non sono d'accordo: 
%% secondo me voi avete assunto che fossero
%% universali, non lo avete ``controllato''
%, and a direct crosscheck at leading twist has been
%performed for two-hadron fragmentation functions in the case of $e^+$-$e^-$
%annihilation~\cite{Boer:2003ya} 
Moreover, when integrating $\Delta^{[-]}(z,\vec k_\st ,R)$ of Eq.~(\ref{eq:linkDeltaz})
upon $\de\vec k_\st$, the displacement of the quark fields is confined to the light-cone
$+$ direction. Hence, the two gauge links $U^T$ and $U^+$ will merge into a single
operator connecting the points $0$ and $\xi$ along a straight line: 
%% qui ho aggiunto un pezzo che indica come la \Delta integrata sia la stessa
%% sia che si parta con il [-] che con il [+]. Se non ti piace puoi rimuoverlo
\begin{equation} \begin{split} 
  \Delta (z, R) &= z^2 \int d\vec k_\st \; \Delta^{[-]} (z,\vec k_\st ,R) 
\;=\; z^2 \int d\vec k_\st \; \Delta^{[+]} (z,\vec k_\st ,R) 
\\ 
 &= z^2 \sumint \; 
  \int \frac{\de \xi^+}{2\pi} \; e^{\ii k\cdot\xi}\;
  \langle 0|\, U^+_{[0,\xi]} \, \psi(\xi) \,|P_h,R; X\rangle
  \langle X; P_h,R|\, \overline{\psi}(0)\,|0\rangle 
  \Big|_{\xi^- = \vec \xi^{}_\st  =0} \; .
  \label{eq:linkDeltak}
\end{split} \end{equation}  
Therefore, in principle $\vec k^{}_\st$-integrated functions are insensitive to the link 
path and should represent universal functions. This is certainly true at leading twist, but
it is still matter of debate at the subleading level because of the presence of 
$\vec k^{}_\st$-weighted contributions~\cite{Boer:2003xz}, as we shall see below. Therefore, here
in the following we will omit an explicit dependence on the gauge link path for
transverse-momentum integrated quantities only when the issue is settled and commonly accepted. 
%% Ho deciso di provare a mettere gli indici dove necessario, cioe` nei pezzi a 
%% subleading twist. Segui le correzioni e vedi se sei d'accordo.
%In the following, since we will be 
%concerned only with $\vec k^{}_\st$-integrated functions, we will assume the general
%attitude about this objects being independent from the gauge link direction and we will 
%omit the $[+]$ and $[-]$ superscripts. 

At subleading twist, both the combinations of the transverse components of the quark 
propagator with $A^-$, and of the $n_+$ projection of the quark propagator with 
$\vec A^{}_\st$, generate a color gauge invariant operator involving the field strength 
tensor $G^{\mu \nu}$. After the $\vec k^{}_\st$-integration, this correlator
reads
%% Su questo ho messo il [-], ho tolto PV e ho rimesso il polo
\begin{equation} \begin{split}
\Delta_A^{[-]\alpha} (z,R) &=  \int \de z_1 \frac{\ii}{z_1 + \ii \epsilon} \Delta_G^\alpha 
(z, z_1, R) \\[1pt]
&= \int \de z_1 \frac{\ii}{z_1 + \ii \epsilon} \, z^2 \sumint \int \frac{\de \xi^+}{2\pi} \, 
\frac{\de \eta^+}{2\pi} \, e^{\ii k\cdot \xi} \, e^{\ii k_1 \cdot (\eta-\xi) } \\
&\qquad \langle 0 | \, U^+_{[0,\xi]} \, \psi (\xi) \, U^+_{[\xi,\eta]} \, g G^{- \alpha} 
(\eta) \, |P_h, R; X \rangle \langle X; P_h, R| \, {\overline \psi} (0) \, | 0 \rangle
\Big\vert_{\xi^- =\eta^-= \vec \xi^{}_\st  =\vec \eta^{}_\st=0} \; ,
\label{eq:defDeltaAt}
\end{split} \end{equation}
where 
%PV indicates the principal value of the integral, after the subtraction of the pole, and 
$z_1=  P_h^- / k_1^-$. This correlator starts contributing at twist 3; it is considered a 
tree-level contribution, as already explained at the beginning of this Section. Again, by 
absorbing the gauge link $U^+$ in a redefinition of the quark field $\psi$, the well known 
expression of the quark-gluon-quark correlator is recovered~\cite{Ellis:1983cd,Jaffe:1992ra}. 
Introducing the covariant derivative $\ii D^{\mu}(\xi) =  \ii \partial^{\mu} + g A^{\mu}(\xi)$ we 
can recover also the relation
%% Qui il [-] va solo su \Delta_A e \Delta_\partial
\begin{equation} 
\Delta_A^{[-]\alpha} (z,R) = \Delta_D^\alpha (z,R) - \Delta^{[-]\alpha}_\partial (z,R) \; ,
\label{eq:linkDeltaA}
\end{equation} 
where
%% Qui ho tolto la media in \Delta_\partial e lasciato solo il [-]
\begin{align} 
\Delta_D^\alpha (z,R) &=  z^2 \sumint \int \frac{\de \xi^+}{2\pi} \, 
e^{\ii k\cdot \xi} \, \langle 0 | \, U^+_{[0,\xi]} \, \psi (\xi) \, \ii D_{}^\alpha (\xi) 
\, |P_h, R; X \rangle \langle X; P_h, R| \, {\overline \psi} (0) \, | 0 \rangle 
\Big\vert_{\xi^- = \vec \xi^{}_\st  =0}, 
\label{eq:linkDeltaD} \\
\Delta^{[-]\alpha}_\partial (z,R) &= z^2 \int \de \vec k^{}_\st \,
k_\st^\alpha \, 
%\half \left( 
%\Delta^{[+]} (z, \vec k^{}_\st, R) + 
\Delta^{[-]} (z, \vec k^{}_\st, R) 
%\right) 
\; , 
\label{eq:linkDeltad}
\end{align} 
with the covariant derivative $D(\xi)$ acting on the left on the quark field $\psi(\xi)$. Note that,
after integrating upon $\vec k^{}_\st$, the term $\Delta_D^\alpha$ becomes insensitive to the gauge
link path. It is possible to relate the quark-gluon-quark correlator to the quark-quark one 
using the equation of motion of QCD, $(\ii \Dslash - m) \psi = 0$. Therefore, 
$\Delta_A^{[-] \alpha} (z,R)$ does not introduce any new fragmentation functions, but it turns 
out that it plays an essential role in ensuring electromagnetic gauge invariance up to 
subleading twist.

In the following, both the quark-quark and quark-gluon-quark correlators of
Eqs.~(\ref{eq:linkDeltaz}) and (\ref{eq:defDeltaAt}), respectively, will be 
analyzed in detail for the semi-inclusive two-hadron production, including the expansion in
the partial waves of the pair.

%%%%%%%%%%%%%%%%%%%%%%%%%%%%%%%%%%%%%%%%%%%%%%%%%%%%%%%%%%%%%%%%%%%%%%%%%%%%%%%%

\subsection{The correlator $\Delta$}
\label{sec:twist2correlator}

The most general parametrization of $\Delta^{[\pm ]} (k,P_h,R)$ in 
Eq.~(\ref{eq:defDelta}), compatible with Hermiticity and parity invariance, is 
given by
%% Sono indeciso se mettere il [-] anche in questa equazione. La
%% decomposizione e` la stessa sia per \Delta^[+] che [-], cambierebbero i
%% coefficienti C_i
\begin{equation}\begin{split} 
\Delta^{[\pm ]} (k,P_h,R)  &=
                M_h\,C_1^{[\pm ]}\,1 
	+ C_2^{[\pm ]}\,\Pslash_h + C_3^{[\pm ]}\,\Rslash + C_4^{[\pm ]}\,\kslash  \\ 
&\quad
 	+ \frac{C_5^{[\pm ]}}{M_h}\,\sigma_{\mu \nu} P_h^{\mspace{2mu}\mu} k^{\nu}
	+ \frac{C_6^{[\pm ]}}{M_h}\,\sigma_{\mu \nu} R^{\mu} k^{\nu}
	+ \frac{C_7^{[\pm ]}}{M_h}\,\sigma_{\mu \nu} P_h^{\mspace{2mu}\mu} R^{\nu} \\ 
&\quad	
        + \frac{C_8^{[\pm ]}}{M_h^2}\,\g_5 \eps^{\mspace{2mu}\mu \nu \rho \sigma}
		\gamma_\mu P_{h \nu} R_\rho k_\sigma \; ,
\label{eq:decomDelta}  
\end{split} 
\end{equation}
where the coefficients $C_i^{[\pm ]}$ are real scalar functions of all the possible
independent invariants, namely $k^2, \, k\cdot P_h, \, k\cdot R, \, M_h^2, \, M_1^2, \, M_2^2$.
Integrating Eq.~(\ref{eq:decomDelta}) upon the suppressed $k^+$ direction and, consequently, 
taking the light-like separation $\xi^-=0$, we get for, e.g., the DIS 
process the following decomposition:
%% Qui sono indeciso se sia il caso di lasciare il [-] o no. In fondo, alla
%%fine non ne abbiamo bisogno perche` sara` comunque integrato su k_T, pero` per 
%%esattezza sarebbe meglio lasciarlo
\begin{equation}\begin{split} 
\Delta^{[-]} (z,  \vec{k}_T, R) 
&= \frac{1}{32 \pi z} \int \de k^+ \Delta^{[-]}(k,P_h,R)\Big|_{k^- = P_h^-/z} 
\\ 
&= \frac{1}{16 \pi}\, \biggl\{ D_1 \,\nslash_- +
H_1^{\open \, \prime}\, \frac{\ii}{2 M_h}\,\left[\Rslash^{}_\st,\, \nslash_-\right] \\ 
&\quad + H_1^{\perp}\, \frac{\ii}{2 M_h}\,\left[\kslash^{}_\st,\, \nslash_-\right] + 
G_1^{\perp} \, \g_5 \frac{\eps_\st^{\rho \sigma} R^{}_{\st \rho} k^{}_{\st
\sigma}}{M_h^2}\, \nslash_-\biggr\} \\ 
& \quad +\frac{M_h}{16 \pi \,P_h^-}\,\biggl\{ E + 
D^{\open \, \prime} \,\frac{\Rslash^{}_\st}{M_h} + D^{\perp} \, 
\frac{\kslash^{}_\st}{M_h} \\ 
& \quad + H\, \frac{\ii}{2}\,\left[\nslash_-,\,\nslash_+ \right] + 
H^{\open}\, \frac{\ii}{2 M_h^2}\,\left[\Rslash^{}_\st,\, \kslash^{}_\st\right] \\ 
& \quad + G^{\open \, \prime} \,\g_5\, 
\frac{\eps_\st^{\rho \sigma} \g_{\rho} R^{}_{\st \sigma}}{M_h} + 
G^{\perp} \,\g_5\, \frac{\eps_\st^{\rho \sigma} \g_{\rho} k^{}_{\st \sigma}}{M_h} 
\biggr\} \; .
\end{split} 
\label{eq:decomFFDelta}
\end{equation}

The first group of terms inside braces represents the leading-twist contribution and
includes the usual interference fragmentation functions (IFF) discussed 
elsewhere~\cite{Bianconi:1999cd,Radici:2001na,Bacchetta:2002ux}. They can be 
obtained by projecting out of Eq.~(\ref{eq:decomFFDelta}) the usual Dirac
structures $\Gamma = \g^-, \, \g^- \g_5, \, \ii \sigma^{i-} \g_5$, where $i$ means a
transverse component. The second group shows the $1/P_h^- \sim 1/Q$-suppressed
fragmentation functions that arise from the Dirac structures $\Gamma = 1, \, \g^i, \, 
\sigma^{-+}, \, \sigma^{ij}, \, \g^i \g_5$, respectively. Note that the structures $\Gamma
= \ii \g_5, \, \sigma^{i+}$, give no contribution at this level. The functions
$H_1^{\open \, \prime}, \, H_1^\perp, \, G_1^\perp, \, H, \, H^{\open}, \, G^{\open \, \prime}, 
\, G^\perp$, are $T$-odd, while $H_1^{\open \, \prime}, \, H_1^\perp, \, E, \, H, \, H^{\open}$, 
are chiral-odd. 

Because of the constraints imposed by kinematics and by the $k^+$ integration, the
fragmentation functions in Eq.~(\ref{eq:decomFFDelta}) actually depend on five
variables, namely 
$z,\, \zeta, \, M_h^2, \, \vec k_\st^{\, 2}, \, \vec k^{}_\st \cdot \vec R^{}_\st$,
and they can generally be decomposed as 
\begin{equation}\begin{split} 
D_1 \bigl(z, \zeta, M_h^2, \vec k_\st^{\, 2}, \vec k^{}_\st \cdot \vec R^{}_\st \bigr) =
D_1^e \bigl(z, \zeta, M_h^2, \vec k_\st^{\, 2}, (\vec k^{}_\st \cdot \vec R^{}_\st)^2 
\bigr) + \frac{\vec k^{}_\st \cdot \vec R^{}_\st}{M_h^2} \, D_1^o \bigl(z, \zeta, M_h^2, 
\vec k_\st^{\, 2}, (\vec k^{}_\st \cdot \vec R^{}_\st)^2 \bigr) \; ,
\end{split} 
\label{eq:FFsplit}
\end{equation}
and similarly for the other functions. Both $D_1^e, D_1^o,$ are even functions of
$\vec k^{}_\st$. 

By integrating Eq.~(\ref{eq:decomFFDelta}) upon the transverse momentum $\vec k^{}_\st$, we
get
%% Idem come sopra
\begin{equation}\begin{split} 
\Delta\bigl(z, R) &\equiv z^2 \int \de \vec k^{}_\st \, 
\Delta^{[-]} (z, \vec{k}_T, R) \\
&= \frac{1}{16 \pi}\, \biggl\{ D_1 \,\nslash_- +
H_1^{\open}\, \frac{\ii}{2 M_h}\,\left[\Rslash^{}_\st,\, \nslash_-\right] \biggr\} \\ 
& \quad +\frac{M_h}{16 \pi \,P_h^-}\,\biggl\{ E + 
D^{\open} \,\frac{\Rslash^{}_\st}{M_h} + 
H\, \frac{\ii}{2}\,\left[\nslash_-,\,\nslash_+ \right] + 
G^{\open} \,\g_5\, \frac{\eps_\st^{\rho \sigma} \g_{\rho} R^{}_{\st \sigma}}{M_h} 
\biggr\} \\
&\equiv \Delta_1\bigl(z, \zeta, M_h^2, \phi_R) + \Delta_2\bigl(z, \zeta, M_h^2, 
\phi_R) \; ,
\end{split} 
\label{eq:decomFFDel}
\end{equation}
where
\begin{equation} \begin{split} 
H_1^{\open} &\equiv H_1^{\open \, \prime e} + H_1^{\perp o (1)} \; , \\
D^{\open} &\equiv D^{\open \, \prime e} + D^{\perp o (1)} \; , \\
G^{\open} &\equiv G^{\open \, \prime e} + G^{\perp o (1)} \; ,
\end{split}
\label{eq:FFeo}
\end{equation}
and each term now depends on $z, \, \zeta, \, M_h^2$. We define the moment of a
fragmentation function as
\begin{equation}
H_1^{\perp\, (1)}(z,\zeta,M_h^2) = \int \de\vec k^{}_\st \, \frac{\vec k_\st^{\,
2}}{2M_h^2}\, H_1^\perp (z,\zeta, M_h^2, \vec k_\st^{\, 2}, \vec k^{}_\st \cdot \vec
R^{}_\st) \; ,
\label{eq:ktmoment}
\end{equation}
and similarly for the other fragmentation functions. The resulting functions 
$H_1^{\open}, \, H, \, G^{\open}$, are still $T$-odd, while $H_1^{\open}, \, E, \, H,$ are
chiral-odd. For later convenience, the leading-twist contribution is indicated by
$\Delta_1$ and the subleading-twist one by $\Delta_2$, respectively.

%%%%%%%%%%%%%%%%%%%%%%%%%%%%%%%%%%%%%%%%%%%%%%%%%%%%%%%%%%%%%%%%%%%%%%%%%%%%%%%%

\subsection{The subleading-twist correlator $\Delta_A^{[-] \alpha}$}
\label{sec:twist3correlator}

As already anticipated above, the color gauge invariant correlator $\Delta_A^{[-] \alpha}$ 
of Eq.~(\ref{eq:defDeltaAt}) is suppressed by one power of $1/Q$
with respect to the leading twist $\Delta^{[-]}$ of Eq.~(\ref{eq:linkDeltaz}). 
Therefore, it must be consistently included when extending the analysis to the 
subleading twist. For sake of simplicity, only the $\vec k^{}_\st$-integrated result
will be shown.
% and, consistently, the link path label will be suppressed
Since the gauge links can be absorbed in a redefinition of the quark fields, both
$\Delta_D^\alpha$ and $\Delta_\partial^{[-] \alpha}$ in Eqs.~(\ref{eq:linkDeltaD}) and
(\ref{eq:linkDeltad}), respectively, can be worked out in a way similar to the 
%the non color gauge invariant 
one-hadron emission. By projecting out 
the usual Dirac structures $\Gamma = \g^-, \, \g^- \g_5, \, \ii \sigma^{i -}\g_5,$ the 
following decomposition results,
%% Qui il [-] ci vuole
\begin{equation}\begin{split} 
\Delta_A^{[-] \alpha}(z, R) &= \Delta_D^{\alpha} \bigl( z,\zeta, M_h^2, \phi_R \bigr) - 
\Delta^{[-] \alpha}_\partial \bigl(z, \zeta, M_h^2, \phi_R \bigr) \\ 
& = \frac{M_h}{16 \pi z}\, \biggl\{ \widetilde{D}^{\open}\, 
\frac{R_\st^{\alpha}}{M_h }\,\nslash_- + \widetilde{G}^{\open}\, 
\frac{\eps_\st^{\alpha \beta} R^{}_{\st \beta}}{M_h}\, \g_5 \nslash_- \\ 
& \quad -\left(\widetilde{E} - \ii\, \widetilde{H} \right)\, 
\frac{\g^{\alpha}\,\nslash_-}{2} - \ii H_1^{\open o (1)}\, 
\frac{R_\st^{\alpha} \Rslash^{}_\st}{M_h^2 }\,\nslash_-  \biggr\} \; ,
\label{eq:decomFFDel3}
\end{split} 
\end{equation}
where the functions with tilde denote
\begin{equation} \begin{split} 
\widetilde{D}^{\open} &\equiv D^{\open} - z\, D_1^{o(1)} \; , \\
\widetilde{G}^{\open}&\equiv G^{\open} - z\, G_1^{\perp (1)}- z\, \frac{m}{M_h}\,
H_1^{\open} \; , \\
\widetilde{E}&\equiv E - z\, \frac{m}{M_h}\, D_1 \; , \\
\widetilde{H}&\equiv H + 2z\, H_1^{\perp (1)} \; , 
\end{split}
\label{eq:tildetw3}
\end{equation} 
and are pure twist-3 fragmentation functions depending on $z, \, \zeta, \, M_h^2$. They all 
vanish in the Wandzura-Wilzcek approximation.

%%%%%%%%%%%%%%%%%%%%%%%%%%%%%%%%%%%%%%%%%%%%%%%%%%%%%%%%%%%%%%%%%%%%%%%%%%%%%%%%

\subsection{Partial-wave expansion}
\label{sec:partial}

If the invariant mass $M_h$ is not very large, the hadron pair can be assumed to be in a
channel corresponding to a relative $s$- or $p$-wave. Consequently, two-hadron 
fragmentation functions can be decomposed in partial waves~\cite{Bacchetta:2002ux}. 
In the center-of-mass (cm) frame of the two hadrons, the emission occurs back-to-back and
the key variable is the angle $\theta$ between the directions of the emission and of $P_h$.
The kinematics described in Sec.~\ref{sec:kin} can be easily adjusted to the cm frame of
the two hadrons; the most important modifications are
\begin{equation} \begin{split}
\vec R^{}_{\tperpb} &= \vec R \sin \theta \; , \\
|\vec R| &= \frac{1}{2M_h} \, \sqrt{M_h^2 - 2(M_1^2+M_2^2) + (M_1^2-M_2^2)^2} \; , \\[2mm]
\zeta &= \frac{1}{M_h} \, \left( \sqrt{M_1^2 - |\vec R|^2} - \sqrt{M_2^2 - |\vec R|^2} - 2
|\vec R| \cos \theta \right) \; ,
\end{split}
\label{eq:theta}
\end{equation}
where the crucial remark is that $\zeta$ is at most a linear polinomial in $\cos \theta$
with coefficients that depend only on invariant masses. This suggests that the dependence
upon $\zeta$ in the fragmentation functions should be conveniently replaced by an expansion
in the Legendre polynomials in $\cos \theta$ and, consequently, the cross section kept
differential in $\de\cos \theta$. The Jacobian $\de \zeta / \de\cos \theta = 2 
|\vec R| / M_h$ can be absorbed in a redefinition of the fragmentation functions. 

The partial-wave expansion of the leading-twist fragmentation functions has been given in
Ref.~\cite{Bacchetta:2002ux}, namely~\footnote{At variance with 
Ref.~\cite{Bacchetta:2002ux}, here we use lower-case indices for the polarization of the
relative partial wave, in order to avoid confusion with the polarization state of the beam
and/or the target in the expression of the cross section (see the following
Sec.~\ref{sec:cross}).}
\begin{equation} \begin{split} 
D_1^{} &\rightarrow D_{1,oo}^{} + D_{1,ol}^{}\cos\theta + 
D_{1,ll}^{} \frac{1}{4}\,(3\cos^2\theta -1) \; , \\
H_1^{\open} &\rightarrow  H_{1,ot}^{\open} + H_{1,lt}^{\open} \cos\theta \; . 
\end{split}
\label{eq:twist2pw}
\end{equation}
Extending the analysis to the subleading-twist functions is straightforward:
\begin{align} 
H_1^{\open \, o \, (1)} &\rightarrow  H_{1,ot}^{\open \, o \, (1)} +
H_{1,lt}^{\open \, o \, (1)} \cos\theta \; , \label{eq:h1mompw} \\ 
\widetilde{D}^{\open} &\rightarrow \widetilde{D}_{ot}^{\open} + \widetilde{D}_{lt}^{\open} 
\cos\theta \; , \label{eq:dopenpw} \\ 
\widetilde{G}^{\open} &\rightarrow \widetilde{G}_{ot}^{\open} + 
\widetilde{G}_{lt}^{\open} \cos\theta \; , \label{eq:gopenpw} \\  
\widetilde{H} &\rightarrow \widetilde{H}_{oo}^{} + \widetilde{H}_{ol}^{}\cos\theta + 
\widetilde{H}_{ll}^{} \frac{1}{4}\,(3\cos^2\theta -1) \; , \label{eq:htildepw} \\
\widetilde{E} &\rightarrow \widetilde{E}_{oo}^{} + \widetilde{E}_{ol}^{}\cos\theta + 
\widetilde{E}_{ll}^{} \frac{1}{4}\,(3\cos^2\theta -1) \; .
\label{eq:etildepw}
\end{align}

%%%%%%%%%%%%%%% Fig. 3 %%%%%%%%%%%%%%%%%%%

\begin{figure}[h]
\begin{center}
\includegraphics[height=4.cm, width=5.cm]{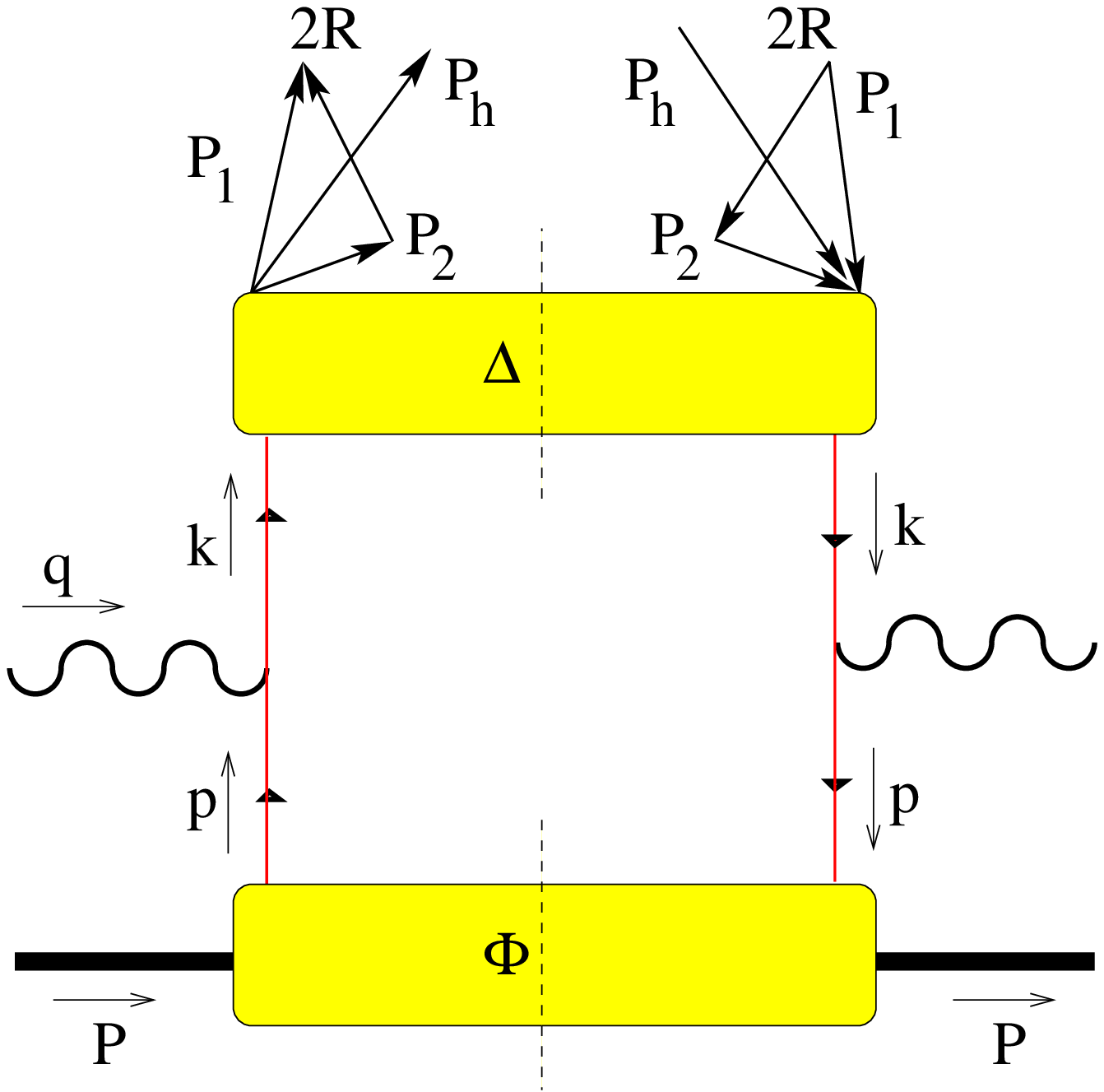}\hspace{0.7truecm}
\includegraphics[height=4.cm, width=5.cm]{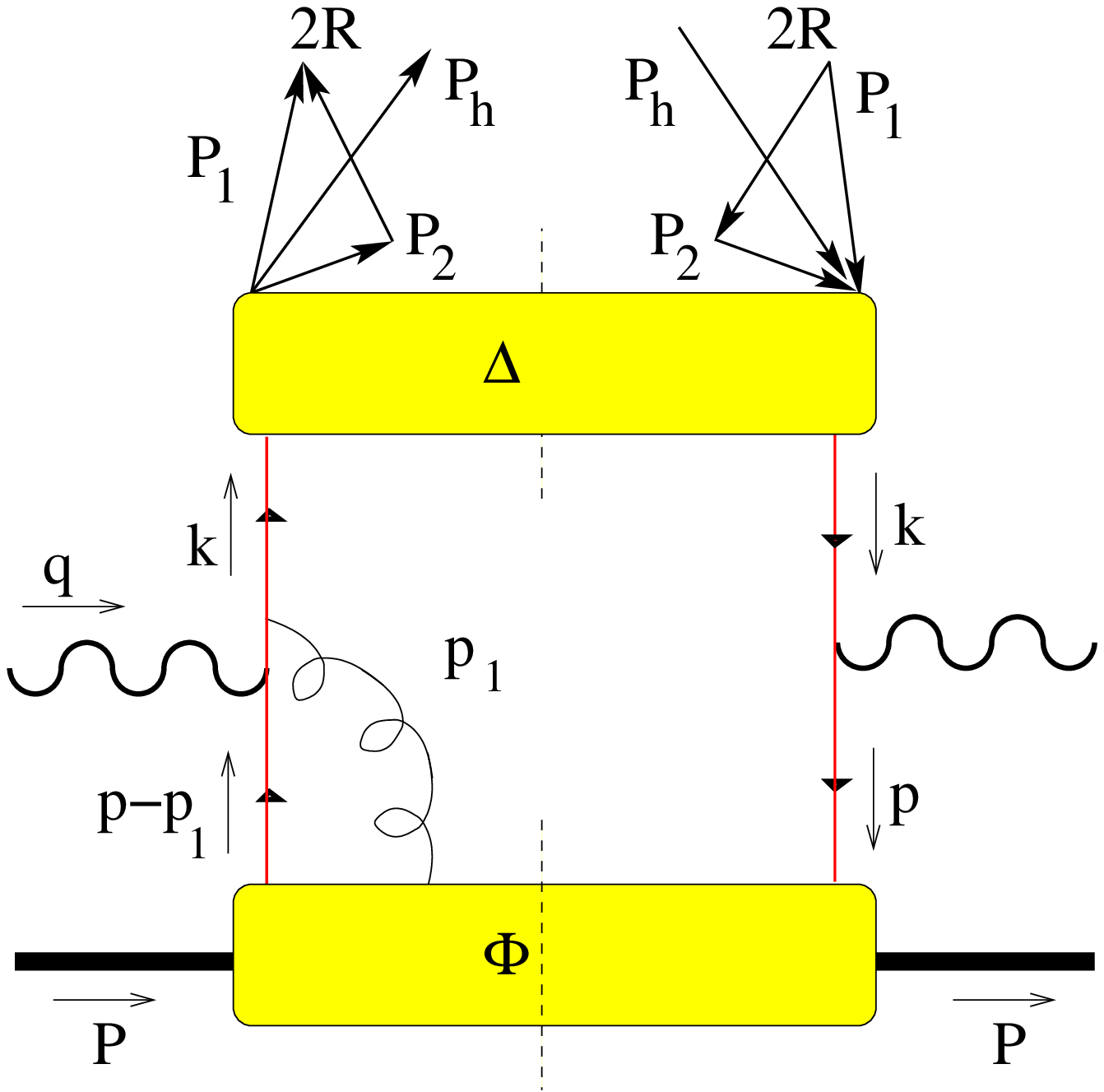}\hspace{0.7truecm}
\includegraphics[height=4.cm, width=5.cm]{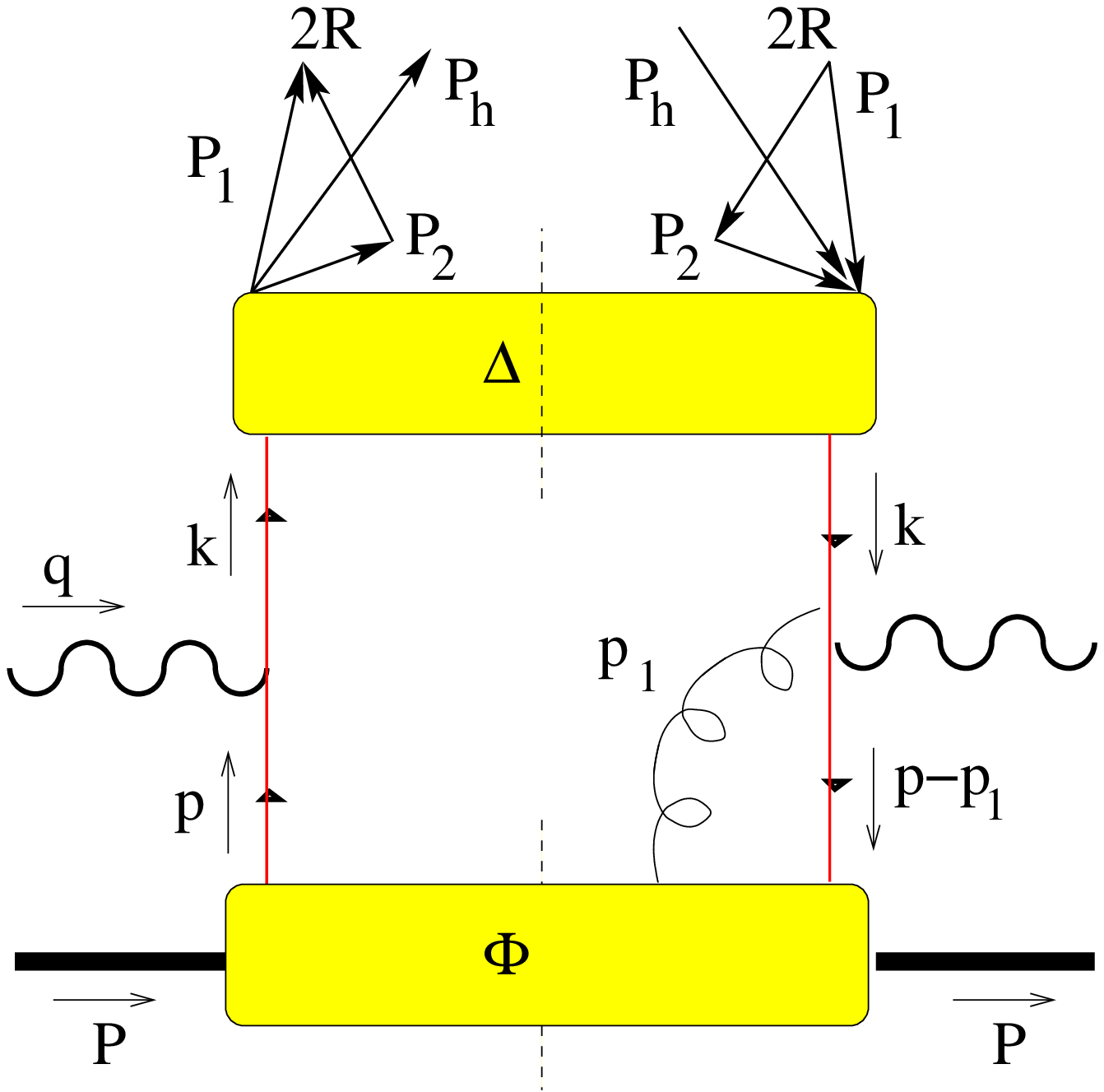}\\[0.3truecm]
{\rm a}\hspace{5.6truecm} {\rm b}\hspace{5.6truecm} {\rm c}\\[0.5truecm] 
\includegraphics[height=4.cm, width=5.cm]{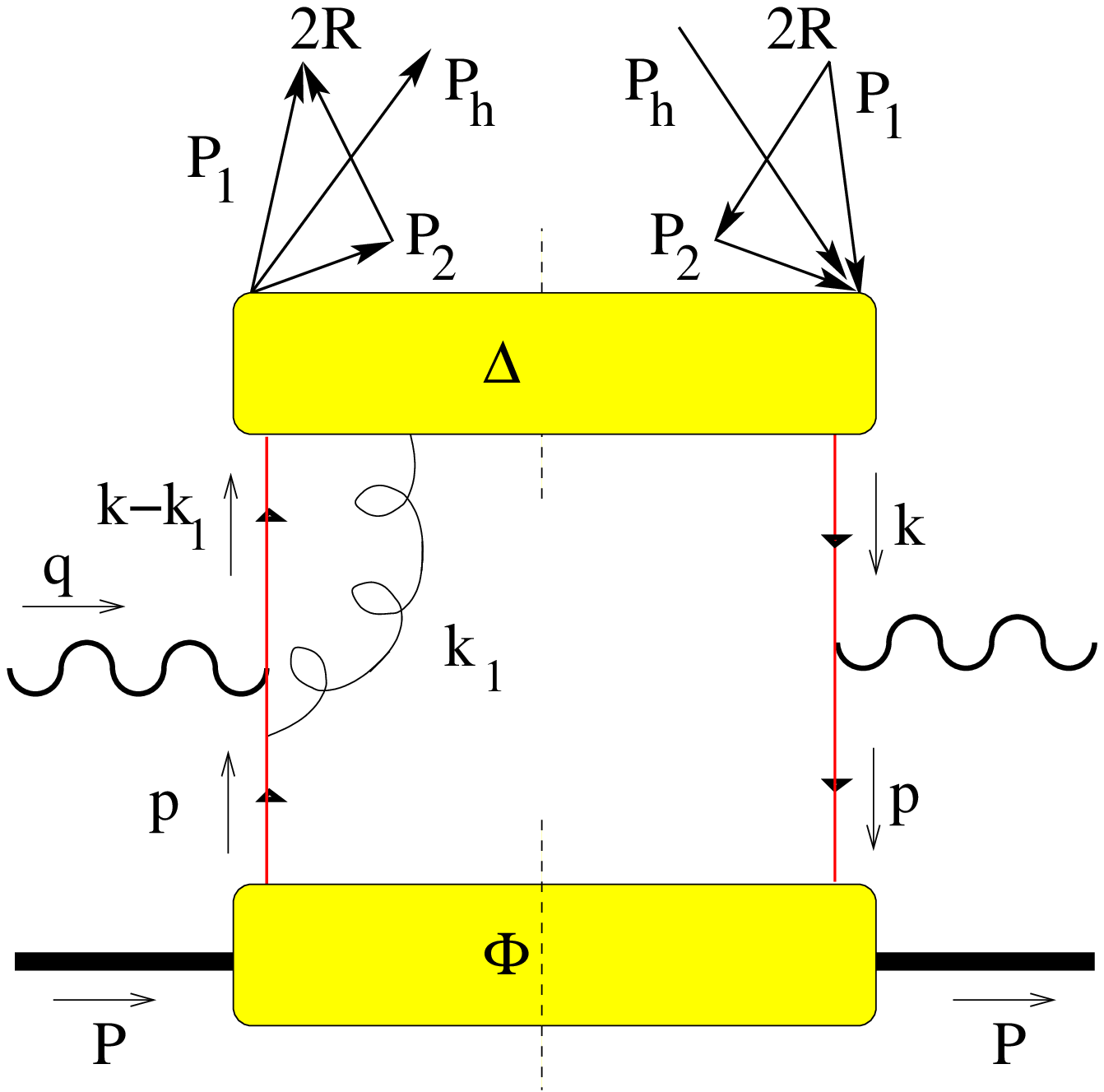}\hspace{0.7truecm}
\includegraphics[height=4.cm, width=5.cm]{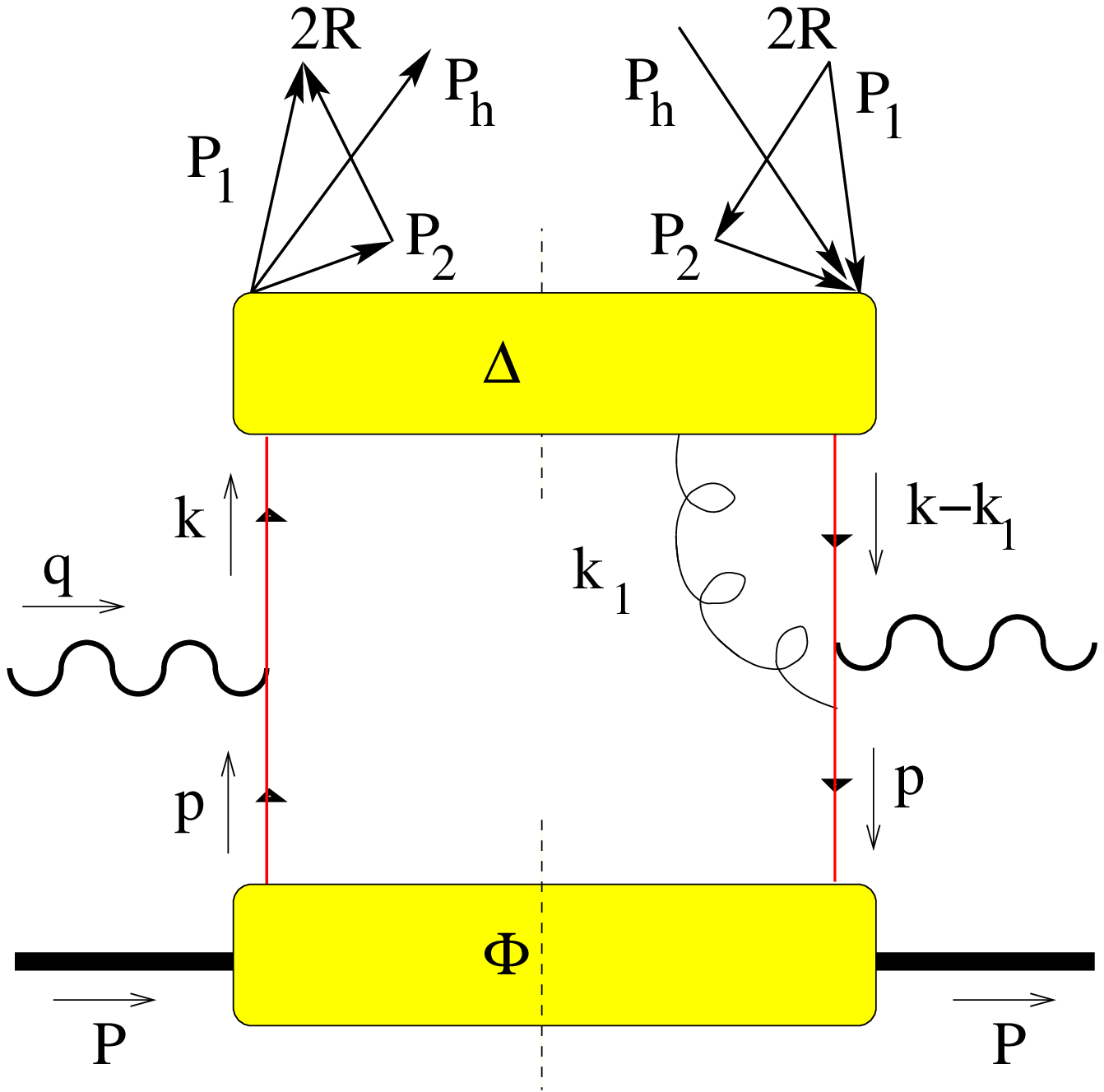}\\[0.3truecm]
{\rm d}\hspace{5.6truecm} {\rm e}
\end{center}
\caption{Relevant diagrams at leading and subleading twist for the SIDIS of a lepton
on a hadronic target with detection of two hadrons in the same current
fragmentation region. The shaded blobs understand the contribution of all
unsuppressed longitudinal gluons, while the gluon lines represent all possible
contributions from transverse gluon fields (see text).}
\label{fig:fig3}
\end{figure}

%%%%%%%%%%%%%%%%%%%%%%%%%%%%%%%%%%%%%%%%%%%%%%%%%%%%%%%%%%%%

%%%%%%%%%%%%%%%%%%%%%%%%%%%%%%%%%%%%%%%%%%%%%%%%%%%%%%%%%%%%%%%%%%%%%%%%%%%%%%%%

\section{Hadronic tensor for semi-inclusive leptoproduction}
\label{sec:hadrontensor}

When the semi-inclusive production of two hadrons happens via a DIS process,
an electron with momentum $l$ scatters off a target nucleon with mass $M$,
polarization $S$ and momentum $P$, via the exchange of a virtual photon with momentum
transfer $q=l-l'$. Inside the target, it is assumed that the photon hits a quark with
momentum $p$, changing it to a state with momentum $k=p+q$ before the fragmentation
(see Fig.~\ref{fig:fig3}a). We define the variable $x=p^+/P^+$, which represents the
light-cone fraction of the target momentum carried by the initial quark. As already
anticipated in Sec.~\ref{sec:kin}, it is customary to consider the frame where all
the hadronic systems have no transverse components, i.e. where $\vec P^{}_\st = \vec
P^{}_{h\,\st}=0$, while the virtual photon has a nonvanishing component 
$\vec q^{}_\st$. A convenient parametrization for the momenta referred to the initial 
hadronic system 
is
\begin{equation} \begin{split} 
   P^\mu &= \left[ \frac{M^2}{2P^+}, P^+, \vec 0 \right]  \\
   p^\mu &= \left[ \frac{p^2+\vec p_\st^{\,2}}{2xP^+}, xP^+, \vec p^{}_\st\right] \; .
\label{eq:kin3}
\end{split} \end{equation}

%%%%%%%%%%%%%%% Fig. 4 %%%%%%%%%%%%%%%%%%%

\begin{figure}[h]
\begin{center}
\includegraphics[width=10.5 cm]{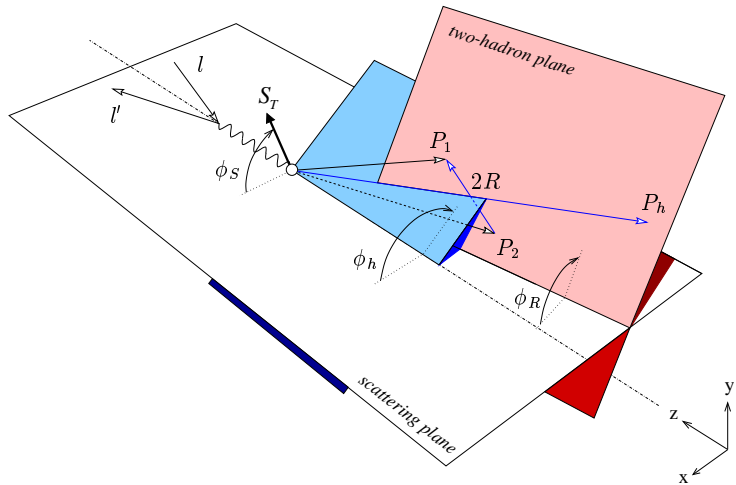}
\end{center}
\caption{Kinematics for the SIDIS of the lepton $l$ on a (un)polarized target leading
to two hadrons inside the same current jet.}
\label{fig:fig4}
\end{figure}
%%%%%%%%%%%%%%%%%%%%%%%%%%%%%%%%%%%%%%%%%%%%

\begin{figure}[h]
\begin{center}
\includegraphics[width=7.cm]{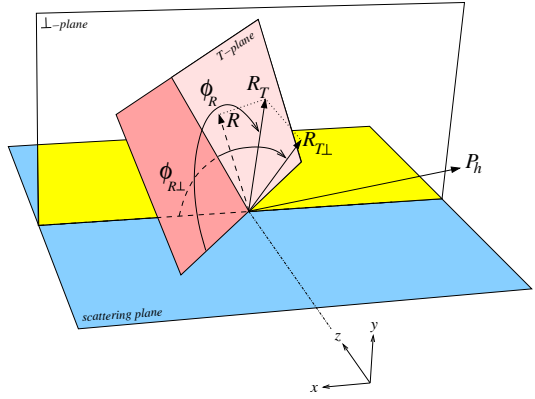}
\end{center}
\caption {Description of the angles $\phi_{R}^{}$ and $\phi_{R \perp}^{}$ .}
\label{fig:fig5}
\end{figure}

However, when calculating the hadronic tensor (and, consequently, the cross section)
it is more convenient to consider the frame where the $\hat z$ axis is antiparallel
to the direction of the virtual photon momentum (see Fig.\ref{fig:fig4}). By 
denoting the momenta in this frame with the subscript $_\perp$, we have, therefore, 
$\vec P_\perp = \vec q_\perp =0$ and $\vec P_{h\,\perp} \simeq -z \vec q^{}_\st$. The
difference between the $_\st$ and the $_\perp$ frames is a boost that introduces
corrections suppressed as $1/Q$; therefore, it can be neglected at leading twist, but
it must consistently be included when extending the analysis at the subleading twist.
The boost amounts to the following modifications, 
\begin{equation} \begin{split} 
n_-^\mu &\sim n_-^{\prime \, \mu} - \frac{\sqrt{2}}{Q} \, q_\st^\mu = n_-^{\prime \,
\mu} + \frac{\sqrt{2}}{Q} \, (p^{}_\st - k^{}_\st )^\mu \; , \\
n_+^\mu &\sim n_+^{\prime \,\mu}  \; , \\
a_\st^\mu &\sim g_\perp^{\mu \nu} a^{}_{\st\,\nu} - \frac{\sqrt{2}}{Q} \, \vec a^{}_\st 
\cdot \vec q^{}_\st \, n_+^\mu \equiv a_{\st\perp}^\mu + \frac{\sqrt{2}}{Q} \, 
\vec a^{}_\st \cdot (\vec p^{}_\st - \vec k^{}_\st) n_+^\mu \; ,
\end{split}
\label{eq:boost}
\end{equation} 
where $a_\st^\mu$ is a generic transverse 4-vector, $n'_\pm$ are the light-like 
vector considered in the $_\perp$ frame, and the analogue of the transverse 
projection operators of Eq.~(\ref{eq:Tproj}) are 
\begin{equation} \begin{split} 
g_\perp^{\mu \nu} &= g^{\mu \nu} - n_+^{\left\{ \mu \right.} n_-^{\left. \prime \,\nu
\right\}} \\
\eps_\perp^{\mu \nu} & \equiv \eps^{\rho \sigma \mu \nu} n_{ + \rho}\, n'_{- \sigma}
\; .
\end{split}
\label{eq:perpproj}
\end{equation} 
As an example of the difference between the $_\st$ and the $_\perp$ frames, 
in Fig.~\ref{fig:fig5} we sketch the vectors $R_T$ and $R_{T\perp}$. 
As expressed in Eq.~(\ref{eq:boost}), the difference between the two vectors is
of order $1/Q$ (exagerated in the drawing). The difference between 
the angles $\phi_{R}^{}$ and $\phi_{R \perp}^{}$ and between $|\vec{R}_\st^{}|$ and
$|\vec{R}_{\tperp}^{}|$ is of order $1/Q^2$, therefore it can be neglected in our 
analysis.

%%%%%%%%%%%%%%% Fig. 5 %%%%%%%%%%%%%%%%%%%

%%%%%%%%%%%%%%%%%%%%%%%%%%%%%%%%%%%%%%%%%%%%%%%%%%%%%%%%%%%%

The hadronic tensor, integrated upon the transverse cm momentum of the hadron
pair, reads
%% Qui ho messo i [-] e i [+] dove necessario. 
\begin{equation}\begin{split} 
2M\,W^{\mu\nu} & = 32 z\, {\rm Tr} \biggl[ z^2 \int \de \vec p^{}_\st \, \de \vec k^{}_\st
\, \Phi^{[+]} (x, \vec p^{}_\st, S)\, \g^{\mu}\, \Delta^{[-]} (z, \vec k^{}_\st, R)\,\g^{\nu}
\biggl] \\ 
& \quad -32 z \, {\rm Tr}\biggl[\g_{\alpha}\, \frac{\g^-}{Q
\sqrt2}\,\g^{\nu}\,\Phi_A^{[+] \alpha}(x, S)\, \g^{\mu}\, \Delta (z, R) \biggr] \\ 
& \quad -32 z \, {\rm Tr}\biggl[\g^{\mu}\, \frac{\g^-}{Q \sqrt2}\,\g_{\alpha}\,
\Delta (z, R)\,\g^{\nu}\,\g^0 \,\Phi_A^{[+] \alpha \dagger}(x,S)
\,\g^0\biggr] \\ 
& \quad -32 z \, {\rm Tr}\biggl[\g^{\nu}\, \frac{\g^+}{Q \sqrt2}\,\g_{\alpha}\,
\Phi(x,S)\,\g^{\mu}\,\g^0 \, \Delta_A^{[-] \alpha \dagger}(z, R) 
\,\g^0\biggr] \\ 
& \quad -32 z \, {\rm Tr}\biggl[\g_{\alpha}\, \frac{\g^+}{Q \sqrt2}\,\g^{\mu}\,
\Delta_A^{[-] \alpha}(z, R)\, \g^{\nu}\,\Phi(x,S) \biggr] \; .
\end{split} 
\label{eq:tensor}
\end{equation}

Each contribution corresponds to a specific class of diagrams in Fig.~\ref{fig:fig3}.
For sake of simplicity, the blobs in the diagrams understand all the connected lines
related to unsuppressed longitudinal gluons, namely the lower blob includes all 
lines with $A^+$ gluons and the upper blob all lines with $A^-$ gluons. Therefore,
the diagram in Fig.~\ref{fig:fig3}a corresponds to the first term in 
Eq.~(\ref{eq:tensor}) involving the leading-twist color gauge invariant correlators 
$\Phi$ (which will be described in Sec.~\ref{sec:phidelta}) and $\Delta$ of 
Eq.~(\ref{eq:linkDeltak}). It is important to perform the integration upon the 
transverse momenta only after including the effect of the boost in
Eq.~(\ref{eq:boost}), as it will turn out that the correlator $\Delta$ contains
a $1/Q$-suppressed $\vec p^{}_T$ dependence. 

At subleading twist, the contribution of transverse gluons $\vec A^{}_\st$ is
symbolically indicated by a line attached to the lower blob, corresponding to the
color gauge invariant quark-gluon-quark correlators $\Phi_A^{[+] \alpha}$ (which will be
described in Sec.~\ref{sec:phidelta}), or to the upper blob, corresponding to 
$\Delta_A^{[-] \alpha}$ in Eq.~(\ref{eq:defDeltaAt}). Therefore, the second and the third 
terms in Eq.~(\ref{eq:tensor}) correspond to diagrams in Figs.~\ref{fig:fig3}b and 
\ref{fig:fig3}c, while the fourth and fifth ones to diagrams in 
Figs.~\ref{fig:fig3}d and \ref{fig:fig3}e, respectively. 

The correlators $\Delta$ and $\Delta^{[-] \alpha}_A$ have already been discussed in
Sec.~\ref{sec:twist2correlator} and \ref{sec:twist3correlator}, respectively. In the following, 
the missing terms will be described in detail, leading to the final 
expression of $W^{\mu \nu}$ in terms of distribution and fragmentation functions.

%%%%%%%%%%%%%%%%%%%%%%%%%%%%%%%%%%%%%%%%%%%%%%%%%%%%%%%%%%%%%%%%%%%%%%%%%%%%%%%%

\subsection{The quark-quark correlators for the initial and the final states}
\label{sec:phidelta}

The color gauge invariant quark-quark correlator for the initial state, $\Phi (x,S)$, 
corresponding to the lower blob in Figs.~\ref{fig:fig3}a, \ref{fig:fig3}d, and 
\ref{fig:fig3}e, reads~\cite{Boer:2003cm}
\begin{equation} \begin{split} 
\Phi (x,S) &= \int \de \vec p^{}_\st \, \Phi^{[+]} (x, \vec p^{}_\st, S) \\
&= \frac{1}{2} \, \biggl\{ f_1(x) \nslash_+ + S_L \, g_1(x) \g_5 \nslash_+ + 
h_1(x) \g_5 \Sslash^{}_{\perp} \, \nslash_+ \biggr\}  \\
&\quad + \frac{\sqrt{2} x M}{2 Q} \, \biggl\{ e(x) + g^{}_\st(x) \g_5 \Sslash^{}_{\perp} + 
S_L \, h_L(x) \g_5 \nslash_+ \nslash_- \biggr\} \\
&\quad + \frac{\sqrt{2} x M}{2 Q} \, \biggl\{ -\ii S_L^{} \, e_L^{} (x) \g_5 - f_\st^{} (x) \, 
\eps_\st^{\alpha \beta} \, \g_\alpha^{} \, S_{\perp \beta}^{} + \ii h(x) \, \nslash_+ 
\nslash_- \biggr\} \\
&\equiv \Phi_1 (x,S) + \Phi_2 (x,S) \; ,
\label{eq:phi}
\end{split} \end{equation}  
where $S_{L/\perp}$ are the longitudinal/transverse components of the target
polarization, respectively. The first group $(\Phi_1)$ represents the contribution of the 
leading-twist distribution functions and it appears in the first, fourth, and fifth, terms of 
Eq.~(\ref{eq:tensor}), corresponding to the diagrams of Fig.~\ref{fig:fig3}a, 
\ref{fig:fig3}d, and \ref{fig:fig3}e, respectively. The other terms $(\Phi_2)$ represent the 
contribution of the subleading-twist distribution functions, including also the $\vec
p_\st^{}$-integrated $T$-odd functions $h(x), \, f_\st^{}(x),$ and $e_L^{}(x)$, which are 
vanishing if the gauge link is the only source of the $T$-odd behaviour. The $\Phi_2$ 
contributes only to the first term of Eq.~(\ref{eq:tensor}), corresponding to the diagram 
of Fig.~\ref{fig:fig3}a. Since the non-integrated $\Phi^{[+]} (x,\vec p^{}_\st, S)$ 
involves scalar products of transverse vectors and commutators between transverse vectors 
and the light-like vector $n_+$, it is easy to check that the boost transformations in 
Eq.~(\ref{eq:boost}) do not add other subleading-twist terms and leave $\Phi (x,S)$ 
unaltered. Moreover, because of the $\vec p^{}_\st$-integration the latter is 
insensitive to the direction of the link integration path.

At subleading twist, also the quark-gluon-quark correlator $\Phi_A^{[+] \alpha}$ comes 
into play, appearing in the diagrams of Figs.~\ref{fig:fig3}b and \ref{fig:fig3}c. 
Similarly to the previous case, the redefinition of the quark fields including the 
gauge links and the integration upon $\vec p^{}_\st$ allows to keep the same relation 
$\Phi_A^{[+] \alpha} = \Phi_D^\alpha - \Phi_\partial^{[+] \alpha}$ dictated by the QCD equations 
of motion as in the non color gauge invariant case. The quark-gluon-quark 
correlator can be parametrized as~\cite{Boer:2003cm}
%% Qui ho aggiunto il [+]
\begin{equation} \begin{split} 
\Phi_A^{[+] \alpha} (x,S) &= \frac{M}{2} \, \biggl\{ \left( x g^{}_\st (x) - \frac{m}{M} \,
h_1(x) - g_{1\st}^{(1)}(x) \right) \, S_{\perp}^\alpha \, \g_5 \nslash_+  \\
& \quad + S_L \, \left( x h_L^{}(x) -\frac{m}{M}\, g_1(x) + 2 
h_{1L}^{\perp\,(1)}(x) + \ii x e_L^{}(x) \right) \, \half \, \g_5 \g^\alpha \nslash_+ \\
& \quad + \left( \frac{m}{M}\, f_1(x) - x e(x) + 2 \ii h_1^{\perp (1)}(x) + \ii x h(x) 
\right) \, \half \, \g^\alpha \nslash_+ \\
& \quad - \left( f_{1\st}^{\perp (1)}(x) + x f_\st^{}(x) \right) \, \eps_\st^{\alpha \beta} \, 
S_{\perp \beta} \, \nslash_+ \biggr\} \; ,
\label{eq:phiA}
\end{split} \end{equation}  
where, as usual, we define the moments
\begin{equation}
g_{1\st}^{(1)}(x) = \int \de\vec p^{}_\st \, \frac{\vec p_\st^{\, 2}}{2M^2} \, 
g^{}_{1\st}(x,\vec p_\st^{\, 2}) \; ,
\label{eq:ptmoment}
\end{equation}
and similarly for the other distribution functions. 

As for the fragmentation into two hadrons, the leading-twist correlator
$\Delta_1$ of Eq.~(\ref{eq:decomFFDel}) occurs in the diagrams of
Figs.~\ref{fig:fig3}a, \ref{fig:fig3}b, and \ref{fig:fig3}c, while the
subleading-twist $\Delta_2$ occurs together with $\Phi_1$ only in the diagram of
Fig.~\ref{fig:fig3}a. However, the boost transformation~(\ref{eq:boost}) induces two
additional contributions to the correlator of Eq.~(\ref{eq:decomFFDelta}), that are
suppressed as $1/Q$ and, therefore, must be consistently included in the analysis at
subleading twist before performing the integration upon $\vec k^{}_\st$. The final 
$\vec k^{}_\st$-integrated result reads
%% Nella prima equazione va il [-], nella seconda no perche` appaiono solo
%% termini a leading twist, anche se il tutto e` soppresso da un 1/Q
\begin{align} 
\begin{split} 
\widetilde{\Delta}_2^{[-]} (z, R) &= \frac{\sqrt{2} M_h}{16 \pi Q} \, 
\biggl\{ - G_1^{\perp\, (1)} \g_5 \frac{\eps_\st^{\mu \rho} \g_\mu
R^{}_{\st\,\rho}}{M_h} - D_1^{o\,(1)} \frac{\Rslash^{}_\st}{M_h}  \\
& \quad  + \ii H_1^{\open\,o\,(1)}
\frac{\vec R_\st^{\,2}}{M_h^2} \frac{1}{2} [\nslash'_-,\nslash_+] + 
\ii H_1^{\perp\,(1)} \, [\nslash'_-,\nslash_+] \biggr\} \; , \label{eq:tildeDelta2} 
\end{split}\\
p_\st^\alpha \, \widetilde{\Delta}_{2\,\alpha} (z, R) &=
\frac{\sqrt{2}}{16 \pi Q} \, p_\st^\alpha \, \biggl\{ D_1 \g_\alpha + 
\ii H_1^{\open} \, \frac{1}{2M_h} [\Rslash^{}_\st, \g_\alpha] + \ii 
\frac{R^{}_{\st\alpha}}{M_h} \, H_1^{\open} \, \frac{1}{2} [\nslash'_-, \nslash_+] 
\biggr\} \; . 
\label{eq:tildeDelta2a}
\end{align} 
Such contributions appear in the first term of Eq.~(\ref{eq:tensor}), corresponding to 
the diagram of Fig.~\ref{fig:fig3}a. The former couples to $\Phi_1$,
while $\widetilde{\Delta}_{2\,\alpha}$, due to the presence of $p_\st^\alpha$, 
couples to
%% Qui ho aggiunto il [+]
\begin{equation} \begin{split} 
\Phi_\partial^{[+] \alpha} (x,S) &\equiv \int \de \vec p^{}_\st \,p^{\alpha}_\st\,
\Phi^{[+]} (x,\vec p^{}_\st, S) \\ 
&= \frac{M}{2} \, \biggl\{ f_{1T}^{(1)}(x)
\eps_T^{\alpha \beta} S_{T \beta} \nslash_+  +   \,g_{1T}^{(1)}(x)\, S_{\perp}^\alpha \, 
\g_5 \nslash_+  \\
& \quad - S_L \, h_{1L}^{(1)}(x)\, \g_5 \g^\alpha \nslash_+ 
- \ii \,h_{1}^{\perp (1)} \, \g^\alpha \nslash_+
\biggr\} \; .
\label{eq:phipartial}
\end{split} \end{equation}  

To complete the picture about the fragmentation at subleading twist, the 
quark-gluon-quark correlator $\Delta_A^{[-] \alpha}$ of Eq.~(\ref{eq:decomFFDel3}) must be 
included in the fourth and fifth terms of Eq.~(\ref{eq:tensor}), corresponding to
diagrams~\ref{fig:fig3}d and \ref{fig:fig3}e.

%%%%%%%%%%%%%%%%%%%%%%%%%%%%%%%%%%%%%%%%%%%%%%%%%%%%%%%%%%%%%%%%%%%%%%%%%%%%%%%%

\subsection{The hadronic tensor}
\label{sec:wmunu}

Putting together in a consistent way all the contributions discussed above up to the
subleading twist, we get for the hadronic tensor the following expression:
\begin{equation}\begin{split} 
2M\,W^{\mu\nu} & = \frac{16 z}{4 \pi}\,\Biggl[ -g_\perp^{\mu \nu} f_1 \,D_1
+ \ii\,\eps_\perp^{\mu \nu}\,S_L \,g_1\, D_1  - 
\frac{R_{\tperp}^{\{\mu} \eps_{\perp}^{\nu \} \rho} S_{\perp \rho}+
S_{\perp}^{\{\mu} \eps_{\perp}^{\nu \} \rho} R_{\tperp \rho}}{2 M_h}\,
h_1 \,H_1^{\open} \\ 
& \quad + S_L\,\frac{2\,\hat t_{}^{\,\{\mu }\eps_{\perp}^{\nu\} \rho} 
R_{\tperp \rho}}{Q}\, \biggl( \frac{M}{M_h}\,\xbj\,h_L\, H_1^{\open} + g_1 \,
\frac{\widetilde{G}^{\open}}{z} \biggr)  \\
& \quad + \frac{2M_h\,\hat t_{}^{\,\{\mu} \eps_\perp^{\nu\}\rho} S_{\perp \rho}}{Q}
\, \biggl[ h_1\,\biggl( \frac{\widetilde{H}}{z}+\frac{\vec R_{\tperpb}^{\,2}}{M_h^2} 
H_1^{\open \, o \,(1)}\biggr) - \frac{M}{M_h} \, x \, f_\st^{} \, D_1 \biggr] + 
\frac{2\,\hat t_{}^{\,\{\mu} R_{\tperp}^{\nu\}}}{Q} \, \biggl( f_1 \,
\frac{\widetilde{D}^{\open}}{z} + \frac{M}{M_h} \, x \, h \, H_1^{\open} \biggr) \\ 
& \quad + \ii\, S_L\, \frac{2\,\hat t_{}^{\,[ \mu}\eps_{\perp}^{\nu] \rho} R_{\tperp \rho}}{Q}\, 
\biggl( g_1\, \frac{ \widetilde{D}^{\open}}{z} - \frac{M}{M_h} \, x \, e_L^{} \, H_1^{\open}
\biggr) - \ii\,\frac{2\,\hat t_{}^{\,[\mu} R_{\tperp}^{\nu]}}{Q}\,
\biggl( \frac{M}{M_h}\,\xbj \,e\, H_1^{\open} + f_1 \,\frac{\widetilde{G}^{\open}}{z} 
\biggr) \\ 
& \quad +\ii\,\frac{2M\,\hat t_{}^{\,[\mu}\eps_\perp^{\nu]\rho}S_{\perp\rho}}{Q}\,
\biggl(\xbj \,g_T\, D_1 +\frac{M_h}{M}\,h_1 \,\frac{\widetilde{E}}{z} \biggr) \Biggr] 
\; . \end{split} 
\label{eq:tensor2}
\end{equation} 
The leading-twist contribution in the above formula involves color-gauge invariant quantities 
that are independent from the properties of the gauge link; under the hypothesis of 
factorization, it represents a universal response. At subleading twist, the issue is still 
under debate~\cite{Boer:2003cm,Metz:2002iz,Boer:2003xz}. However, it is interesting to note 
that only $\vec k^{}_\st$-integrated fragmentation functions appear via 
$\Delta^{[-] \alpha}_\partial$ in Eq.~(\ref{eq:decomFFDel3}), leading to the ``tilde'' functions 
of Eq.~(\ref{eq:tildetw3}), that might depend on the considered process. No
$\vec p^{}_\st$-integrated distribution functions appear via the corresponding
$\Phi^{[+] \alpha}_\partial$, because these contributions in Eq.~(\ref{eq:phiA}) are exactly 
cancelled by the ones generated by coupling $\widetilde{\Delta}_{2\,\alpha}$ of 
Eq.~(\ref{eq:tildeDelta2a}) to $\Phi^{[+] \alpha}_\partial$ of Eq.~(\ref{eq:phipartial}) in the 
first, second, and third contributions of Eq.~(\ref{eq:tensor}) (see also Eq.~(64) of 
Ref.~\cite{Boer:2003cm}). The net
result is that the functions $h(x), f_\st^{}(x),$ and $e_L^{}(x)$, are the only $T$-odd 
distributions in the hadronic tensor and, if not vanishing, they must be generated by a
dynamical mechanism that has nothing to do with the sensitivity to the link path.

%%%%%%%%%%%%%%%%%%%%%%%%%%%%%%%%%%%%%%%%%%%%%%%%%%%%%%%%%%%%

\section{Cross section and spin asymmetries}
\label{sec:cross}

The cross section for SIDIS of polarized leptons off polarized hadronic targets with 
two unpolarized hadrons in the same current fragmentation region, reads
\begin{equation}
\frac{\de^7\! \sigma}{\de\zeta\,\de M_h^2\,\de\phi_R^{}\,\de z\,\de\xbj\,\de y\,
\de\phi_S^{}} = \sum_a \frac{\alpha^2 y \, e_a^2}{32 z Q^4}\, L_{\mu \nu} \, 
2 M W_a^{\mu \nu} \; , 
\label{eq:cross}
\end{equation}
where $\alpha$ is the fine structure constant, $y=(E-E')/E$ is the fraction of beam 
energy transferred to the hadronic system and it is related to the lepton scattering angle 
in the target rest frame, $\phi_S^{}$ is the azimuthal angle of the target polarization with 
respect to the scattering plane, $\phi_R^{}$ is the azimuthal angle of the 
$\vec R^{}_\st$ vector with respect to the scattering plane, measured either around 
the $P_h$ direction or around the $\hat z$ direction (see Fig.~\ref{fig:fig5}). The 
indicated sum runs over the quark and antiquark flavors $a$. The hadronic tensor 
$W_a^{\mu \nu}$ of Eq.~(\ref{eq:tensor2}) is contracted with the lepton tensor
\begin{eqnarray}
L^{\mu \nu} &= &\frac{Q^2}{y^2} \, \biggl[ -2 A(y) g_\perp^{\mu \nu} + 4 B(y) 
\hat t^\mu \hat t^\nu + 4 B(y) (\hat x^\mu \hat x^\nu + \half g_\perp^{\mu \nu}) + V(y) 
\hat t^{\left\{ \mu \right.} \hat x^{\left. \nu \right\}}  \nn \\
& &\qquad + 2 \ii \lambda C(y) \eps_\perp^{\mu \nu} - \ii \lambda W(y) \hat t_{}^{\left[ 
\mu \right.} \eps_\perp^{\left. \nu \right] \rho} \hat x^{}_\rho \biggr] \; ,
\label{eq:leptontensor}
\end{eqnarray}
where $\lambda$ is the lepton helicity, $\hat x$ the spatial unit vector, 
$\hat t_{}^\mu = (n_+^\mu + n_-^{\prime\,\mu})/\sqrt{2}$, and 
\begin{equation} \begin{split} 
A(y) &= \left(1-y+\frac{y^2}{2}\right) \; , \\
B(y) &=(1-y) \; , \\ 
C(y) &= y\left( \frac{y}{2}-1\right) \; , \\
V(y) &= 2\,(2-y)\, \sqrt{1-y} \; , \\
W(y) &= 2\, y \,  \sqrt{1-y} \; .
\end{split}
\label{eq:leptondefs}
\end{equation} 

For convenience, in the following we will indicate the unpolarized or longitudinally
polarized states of the beam with the labels $O$ and $L$, respectively. Similarly, we
will use the labels $O, L, T,$ to indicate an unpolarized, longitudinally polarized, 
transversely polarized, target. We can then deduce the following list of cross 
sections~\footnote{The distribution and fragmentation functions are understood to 
have a flavor index $a$.}:
\begin{widetext}
\begin{align} 
\begin{split}
\de^7\! \sigma^{}_{OO} &= \frac{\alpha^2}{2\pi Q^2 y}\,\sum_a e_a^2 \Biggl\{
     A(y)\,f_1(\xbj)\, D_1\bigl(z,\zeta, M_h^2\bigr)  \\ 
     & \qquad \quad - V(y)\,\cos{\phi_R^{}}\,\frac{|\vec R^{}_{\tperpb}|}{Q}\, \biggl[ 
     \frac{1}{z}\, f_1(\xbj)\,\widetilde{D}^{\open}\bigl( z,\zeta, M_h^2 \bigr) + 
     \frac{M}{M_h} \, x \, h(x) \, H_1^{\open}\bigl( z,\zeta, M_h^2 \bigr) \biggr] \Biggr\} 
     \; ,
\label{eq:crossOO}
\end{split}  \\
\begin{split}
\de^7\! \sigma^{}_{OL} &= \frac{\alpha^2}{2\pi Q^2 y}\, S_L \sum_a e_a^2 \, 
   V(y)\,\sin{\phi_{R}}\, \frac{|\vec R^{}_{\tperpb}|}{Q}\,\biggl[
   \frac{M}{M_h}\,\xbj\, h_L(\xbj)\, H_1^{\open}\bigl(z,\zeta, M_h^2\bigr) +
   \frac{1}{z} \, g_1(\xbj)\,\widetilde{G}^{\open}\bigl(z,\zeta, M_h^2\bigr)\biggr] \; ,
\label{eq:crossOL}
\end{split}  \\
\begin{split}
\de^7\! \sigma^{}_{OT} &= \frac{\alpha^2}{2\pi Q^2 y}\, |\vec S_{\perp}^{}| \sum_a e_a^2
   \,\Biggl\{ B(y)\, \sin(\phi_R^{} + \phi_S^{})\,\frac{|\vec R^{}_{\tperpb}|}{M_h}\,
   h_1(\xbj)\,H_1^{\open}\bigl(z,\zeta, M_h^2\bigr) \\ 
  & \qquad \quad +V(y)\,\sin{\phi_S^{}}\,\frac{M_h}{Q}\,\biggl[ h_1(\xbj)\,\biggl( \frac{1}{z}
  \, \widetilde{H}\bigl(z,\zeta,M_h^2\bigr) +\frac{|\vec R^{}_{\tperpb}|^2}{M_h^2}\,
   H_1^{\open \, o \,(1)}\bigl(z,\zeta,M_h^2\bigr) \biggr) - \frac{M}{M_h} \, x \, f_\st^{}
   (x) \, D_1^{}\bigl(z,\zeta,M_h^2\bigr) \biggr] \Biggr\} \; ,
\label{eq:crossOT}
\end{split} \\
\begin{split}
\de^7\! \sigma^{}_{LO} &=  \frac{\alpha^2}{2\pi Q^2 y}\,\lambda \sum_a e_a^2 \,
    W(y)\,\sin{\phi_{R}}\,\frac{|\vec R^{}_{\tperpb}|}{Q}\,\biggl[
    \frac{M}{M_h}\,\xbj\, e(\xbj)\, H_1^{\open}\bigl(z,\zeta, M_h^2\bigr)
    +\frac{1}{z}\,f_1(\xbj)\,\widetilde{G}^{\open}\bigl(z,\zeta, M_h^2\bigr)\biggr] \; ,
\label{eq:crossLO}
\end{split} \\
\begin{split}
\de^7\! \sigma^{}_{LL} &= \frac{\alpha^2}{2\pi Q^2 y}\,\lambda \, S_L\,\sum_a e_a^2\, 
   \Biggl\{ C(y)\, g_1(\xbj)\, D_1\bigl(z,\zeta, M_h^2\bigr) \\
   &\qquad \quad - W(y)\,\cos{\phi_R^{}}\,\frac{|\vec R^{}_{\tperpb}|}{Q}\,\biggl[ 
   \frac{1}{z}\,g_1(\xbj)\,\widetilde{D}^{\open}\bigl(z,\zeta, M_h^2\bigr)  -
   \frac{M}{M_h} \, x \, e_L^{} (x) \, H_1^{\open}\bigl(z,\zeta, M_h^2\bigr)\biggr] \Biggr\} 
   \; ,
\label{eq:crossLL}
\end{split} \\
\begin{split}
\de^7\! \sigma^{}_{LT} &= \frac{\alpha^2}{2\pi Q^2 y}\,\lambda\,|\vec S_{\perp}|\, 
    \sum_a e_a^2\, W(y)\,\cos{\phi_S}\,\frac{M_h}{Q}\,\biggl[
     -\frac{M}{M_h}\,\xbj\, g_T(\xbj)\,D_1\bigl(z,\zeta, M_h^2\bigr)
     -\frac{1}{z}\,h_1(\xbj)\,\widetilde{E}\bigl(z,\zeta, M_h^2\bigr) \biggr] \; .
\label{eq:crossLT}
\end{split}
\end{align} 
\end{widetext}

In the above formula, we stress again that we have included also the contributions of the $\vec
k_\st^{}$-integrated $T$-odd distribution functions $h(x), f_\st^{}(x),$ and $e_L^{}(x)$,
which are vanishing if the gauge link is the only source of a $T$-odd behaviour. It would
be interesting to experimentally check this feature. 

Several useful spin asymmetries can also be built out of the previous formulae. In 
Eq.~(\ref{eq:crossOT}) for $\de^7 \sigma^{}_{OT}$, the transversity $h_1$ can be isolated 
at leading twist through the fragmentation function $H_1^{\open}$ in a 
$\sin (\phi_R^{}+\phi_S^{})$ spin asymmetry. This asymmetry has been already
discussed in leading-order analyses~\cite{Radici:2001na,Bacchetta:2002ux}
and seems very promising with respect to the Collins asymmetry, since it does not need to 
keep memory of the $\vec k^{}_\st$ dependence but rather of the direction of $\vec R^{}_\st$. 

While data from purely transversely 
polarized targets are not yet available, the HERMES collaboration has performed spin 
asymmetry measurements with targets longitudinally polarized along the lepton
beam~\cite{Airapetian:2000tv,Airapetian:2001eg,Airapetian:2002mf}, 
hence with a polarization 3-vector $\vec S =(S_x,0,S_z)$ in the lepton scattering plane 
$(\phi_S^{}=0)$ and with a transverse component $S_x$ with respect to the direction of the
momentum transfer along $\hat z$. Because of the kinematics setup, $S_x$ is 
suppressed by $1/Q$ with respect to $S_z$~\cite{Oganessyan:1998ma}. In the present case 
of detection of two hadrons in the same jet, therefore, both the leading-twist 
$\de^7 \sigma^{}_{OT}$ and subleading-twist $\de^7 \sigma^{}_{OL}$ of 
Eqs.~(\ref{eq:crossOT}) and (\ref{eq:crossOL}), respectively, should be consistently
considered at the same time when looking for a $\sin\phi_R^{}$ asymmetry. However, 
$\de^7 \sigma^{}_{OT}$ is considerably simpler than the corresponding cross section for
one-hadron SIDIS, because the information about the transversity is not contaminated by
other contributions, as it happens with the Collins and Sivers effects,
respectively. Moreover, in the Wandzura-Wilzcek approximation the fragmentation function 
$\widetilde{G}^{\open}$ vanishes inside $\de^7 \sigma^{}_{OL}$; therefore, 
a $\sin\phi_R^{}$ spin asymmetry for two-hadron SIDIS in the
HERMES kinematics would approximately lead to the product of the fragmentation
function $H_1^{\open}$ times the transversity $h_1$ and the distribution $h_L^{}$, which 
is anyway related to $h_1$ itself via a Wandzura-Wilzcek integral relation.

Again, if we neglect $\widetilde{G}^{\open}$, a $\sin\phi_R^{}$ spin
asymmetry with polarized beam and unpolarized target would give access to the
chiral-odd distribution $e(x)$, always through the chiral-odd fragmentation function 
$H_1^{\open}$, as it is evident from inspection of $\de^7 \sigma^{}_{LO}$ in 
Eq.~(\ref{eq:crossLO}). The function $e(x)$ has recently attracted a lot of 
interest~\cite{Efremov:2002ut}, 
because it is directly related to the soft physics of chiral symmetry
breaking~\cite{Jaffe:1991kp}. Its first isoscalar Mellin moment gives the scalar form 
factor. Although this form factor (describing the elastic scattering off a
spin-$\half$ target via the exchange of a spin-0 particle) has not yet been measured, its
value at $t=-Q^2=0$, the so-called $\sigma$ term, can be deduced by low-energy theorems 
from the experimental pion-nucleon scattering in the time-like region at the so-called
Chen-Dashen point $t=-Q^2=2m_\pi^2$, with $m_\pi$ the pion
mass~\cite{Weinberg:1966,Cheng:1971,Brown:1971}. Unexpectedly, the
$\sigma$ term turns out very big (50-70 MeV)~\cite{Gasser:1991ce,Olsson:2002pi} with 
respect to the average value of available lattice calculations~\cite{Leinweber:2003dg}, 
suggesting that approximately 20\% of the nucleon mass $M$ could be due to the strange 
quark content of the nucleon. Therefore, having experimental access to $e(x)$ is of great 
importance. This distribution could be extracted at subleading twist through the Collins 
function by a beam spin asymmetry in one-hadron SIDIS for longitudinally polarized beams 
and unpolarized targets~\cite{Mulders:1996dh,Avakian:2003pk}, provided that the transverse 
momentum of the detected hadron is measured. This asymmetry contains another contribution 
that was neglected until recently~\cite{Yuan:2003gu,Gamberg:2003pz}. Once again, the case 
of one-hadron SIDIS is complicated by the dependence upon the partonic transverse momentum. 
For the case of two-hadron SIDIS, it is possible to integrate upon the transverse total 
momentum of the pair and still build an azimuthal asymmetry using $\vec R^{}_\st$. In fact, 
Eq.~(\ref{eq:crossLO}) looks simpler than the corresponding one for the one-hadron case, 
and it could eventually represent the cleanest channel to look at in order to extract $e(x)$.

Finally, when expanding the fragmentation functions in partial waves and making the cross
section differential in $\cos\theta$, the different dependence upon $\theta$ allows to 
distinguish the contributions pertaining to pure
$s$ waves, pure $p$ waves, and $s$-$p$ interferences. For instance, by
substituting Eqs.~(\ref{eq:twist2pw}) and (\ref{eq:theta}) into
Eq.~(\ref{eq:crossOT}) it is possible to check that the asymmetry will be
dominated by an $s$-$p$ interference fragmentation function at $\theta= \pi/2$, 
and by a $p$-wave interference fragmentation function at $\theta= \pi/4$.

%%%%%%%%%%%%%%%%%%%%%%%%%%%%%%%%%%%%%%%%%%%%%%%%%%%%%%%%%%%%

\section{Conclusions}
\label{sec:end}

Fragmentation functions are universal, process-independent objects~\cite{Collins:1982uw} 
containing a crucial information about the hadronization mechanism and, ultimately, about 
the confinement of partons inside hadrons. They appear in semi-inclusive processes such as, 
e.g., DIS or electron-positron annihilation, and they can act also as a 
sort of ``analyzing power'' for the polarization state of the fragmenting 
quark~\cite{Efremov:1982sh,Artru:1990zv,Jaffe:1993xb,Ji:1994vw}. The typical example is the 
so-called Collins effect~\cite{Collins:1993kk,Collins:1994kq} relating the transverse 
polarization of the parent quark to the transverse-momentum dependent Collins function, 
that describes a (nonperturbative) azimuthal asymmetry in the distribution of the detected 
leading hadron. Two-hadron fragmentation functions can also be defined, among which the 
so-called interference fragmentation 
functions~\cite{Collins:1994ax,Jaffe:1998hf,Bianconi:1999cd} lead to 
interesting single-spin asymmetries even after integrating upon the transverse total 
momentum of the pair~\cite{Radici:2001na,Boer:2003ya}, thus avoiding the complications 
introduced by the intrinsic nonperturbative transverse-momentum dependence of the Collins 
function. 

In this paper, we have extended the analysis of two-hadron fragmentation
functions to the subleading-twist level, discussing also the issue of color gauge
invariance but eventually integrating upon the transverse total momentum of the pair. Our 
results are theoretically interesting because the absence of an intrinsic
nonperturbative dependence upon transverse momenta cancels, at leading twist, also any 
dependence upon the properties of the gauge link operator necessary to restore gauge 
invariance, allowing for a truly universal definition of these objects; a debate is still
ongoing to check if this property holds true also at subleading
twist~\cite{Metz:2002iz,Boer:2003cm,Boer:2003xz}. 
%% ho aggiunto la references 
The extension to the 
subleading twist is also experimentally important, because it can represent a nonnegligible
contribution when performing measurements at moderate $Q^2$. 

We have analyzed both the quark-quark and the suppressed quark-gluon-quark correlator, 
relating the latter to the former by means ot the QCD equations of motion.
We have presented the full decomposition up to the subleading-twist level of these  
correlators in terms of fragmentation functions integrated upon the 
intrinsic transverse momentum. As previously stressed, these functions are universal certainly
at twist 2 and maybe also at twist 3.  
%% ho cambiato probably con maybe

As an application of our results, we have calculated the hadronic tensor and the cross
section for all possible combinations of polarization states of the beam-target system 
in the case of deep-inelastic semi-inclusive leptoproduction of two unpolarized hadrons, by 
integrating upon the two-hadron center-of-mass transverse momentum.
Our results can be used to distinguish $1/Q$-suppressed contributions in
experimental measurements, in order to extract more clearly leading-twist contributions, 
or in order to study interesting subleading-twist terms. An example of the former case is
the possibility of extracting the transversity distribution in spin asymmetries also with
longitudinally polarized targets (as they have been measured at
HERMES~\cite{Airapetian:2002mf} for the case of one-hadron production); an example of the latter 
is the possibility of extracting from beam-spin asymmetries (probably in the cleanest 
possible way~\cite{Yuan:2003gu,Gamberg:2003pz}) the twist-3 chiral-odd distribution function 
$e(x)$~\cite{Avakian:2003pk}, related to the mechanism of the spontaneous breaking of the 
QCD chiral symmetry and, ultimately, to the strange-quark content of the 
nucleon~\cite{Jaffe:1991kp}.

As a last step, we have performed a partial-wave expansion of leading- and 
subleading-twist two-hadron fragmentation functions, in order to distinguish the 
interference coming from the $s$-$s$, $p$-$p$, and $s$-$p$ channels in the relative 
partial wave of the hadron pairs. Each component carries information on different 
mechanisms, such as the polarization transfer to spin-1 resonances (for $p$-$p$ 
interference) or $T$-odd effects from different kinds of final-state interactions. 
Therefore, extracting this information from data would allow for the exploration 
of different aspects of the physics of the fragmentation process.

%%%%%%%%%%%%%%%%%%%%%%%%%%%%%%%%%%%%%%%%%%%%%%%%%%%%%%%%%%%%%%%%%%%%%%%%%%%%%%%%%%%%

\begin{acknowledgments}
Several discussions with D.~Boer, P.~J.~Mulders, and F.~Pijlman,
are gratefully acknowledged. This work has been partially
supported by the TMR network HPRN-CT-2000-00130 and by the BMBF. 
M.~R. thanks the Institute for Nuclear Theory at the University of 
Washington (Seattle, USA) for its hospitality and the Department of Energy for partial 
support during the completion of this work.
\end{acknowledgments}

%%%%%%%%%%%%%%%%%%%%%%%%%%%%%%%%%%%%%%%%%%%%%%%%%%%%%%%%%%%%%%%%%%%%%%%%%%%%%%%%%%%%%%

\appendix

%%%%%%%%%%%%%%%%%%%%%%%%%%%%%%%%%%%%%%%%%%%%%%%%%%%%%%%%%%%%%%%%%%%%%%%%%%%%%%%%%%%%

%%%%%%%%%%%%%%%%%%%%%%%%%%%%%%%%%%%%%%%%%%%%%%%%%%%%%%%%%%%%%%%%%%%%%%%%%%%%%%%%%%%%%

\bibliographystyle{apsrev}
\bibliography{mybiblio}

\end{document}